\documentclass[mathpazo]{cicp}
\usepackage{CJK}
\usepackage{latexsym,bm}
\usepackage{amssymb}
\usepackage{fancyhdr}
\usepackage{cite}
\usepackage{tabularx}
\usepackage{booktabs}
\usepackage{dcolumn}
\usepackage{graphicx}
\usepackage{subfigure}
\usepackage{wrapfig}
\usepackage{multicol}
\usepackage{verbatim}
\usepackage{mathrsfs}
\usepackage{multirow}
\usepackage{indentfirst} 
\usepackage{graphicx}
\usepackage{diagbox}
\usepackage[section]{placeins}
\usepackage{multirow}

\def\firstpage{1}                           

\newcommand{\supercite}[1]{\!\!\textsuperscript{\cite{#1}}}

\def\geq{\geqslant}

\begin{document}

\begin{CJK*}{GBK}{song}
\title{{\large  \textbf{Continuous and discontinous compressible flows in a converging-diverging channel solved
by physics-informed neural networks without data}}
}
\author[Zhang Z R et.~al.]{Hong Liang\affil{1},Zilong Song\affil{1},
      Chong Zhao\affil{3}, and Xin Bian\affil{2}\comma\corrauth}
 \address{\affilnum{1}\ Department of Physics, Hangzhou Dianzi University, Hangzhou 310018, China. \\
          \affilnum{2}\ State Key Laboratory of Fluid Power and Mechatronic Systems,\\ Department of Engineering Mechanics, Zhejiang University, Hangzhou 310027, China.
          \affilnum{3}\  Hangzhou Shiguangji Intelligient Electronics Technology Co., Ltd, Hangzhou 310018, China.}        
 \emails{{\tt bianx@zju.edu.cn} (X.~Bian)}
\begin{abstract}
Physics-informed neural networks (PINNs) are employed to solve the classical compressible flow problem in a converging-diverging nozzle. This problem represents a typical example described by the Euler equations, 
thorough understanding of which serves as a guide for solving  more general compressible flows. Given a geometry of the channel, analytical solutions for the steady states indeed exist and they depend on the ratio between the back pressure of the outlet and stagnation pressure of the inlet. Moreover, in the diverging region, the solution may branch into subsonic flow, supersonic flow, and a mixture of both with a discontinuous transition where a normal shock takes place. Classical numerical schemes with shock-fitting/capturing methods have been designed to solve this type of problem effectively, whereas the original PINNs fail in front of the hyperbolic non-linear partial differential equations. We make a first attempt to exploit the power of PINNs to directly solve this problem by adjusting the weights of different components of the loss function, to acquire physical solutions and meanwhile avoid trivial solutions. With a universal setting yet no exogenous data,
we are able to solve this problem accurately, that is, for different given pressure ratios PINNs provide different branches of solutions at both steady and unsteady states, some of which are discontinuous in nature.
\end{abstract}

\ams{35L51, 76L05, 76N06}
\keywords{unsteady compressible flow, normal shock, physics-informed neural networks, direct numerical simulation.}
\maketitle

\section{Introduction}
\small Euler equations embrace the conservation laws for inviscid fluids, which are often compressible at high speed~\supercite{white2016}. It is infeasible to derive an analytical solution for this type of equations except in a few special cases. A discontinuous solution associated with shock wave may be generated due to the hyperbolic and non-linear properties of the partial differential equations (PDEs) when the fluid moves at speeds comparable to its speed of sound~\supercite{dafermos2005hyperbolic}.  This further poses challenges for the development of numerical schemes and therefore, have induced many ingenious efforts in the past a few decades~\supercite{anderson1995, leveque2002finite}. There is generally a trade-off between accuracy and stability in the numerical methods~\supercite{shu2013brief}. The low order methods can produce stable but less accurate results, where sharp profiles of shocks are smoothed. To the contrary, the high-order methods are able to generate relatively accurate results, but are often troubled by instabilities and Gibbs phenomenon near the discontinuities~\supercite{jiang1996efficient}. Furthermore, the requirement of stability also imposes a strict CFL limit to the numerical methods, resulting in small time steps in simulations. These facts render the simulations by traditional numerical methods both human-intelligence concentrated and computationally intensive~\supercite{wang2007high,jiang2023gas,yu2018revisit,hou2023high}. 

Recently, machine learning methods, in particular, the deep neural networks (DNNs) have received enormous attentions and are trying to replace human pounding by network training and infering. Due to the tremendous successes in other fields, they also quickly became an alternative approach to solve PDEs~\supercite{michoski2019solving,wang2022active,bezgin2022weno3,liu2019neural}. When solving the Euler equations, DNNs can be embedded into the traditional numerical methods to facilitate accurate results~\supercite{magiera2020constraint,schwander2021controlling,bezgin2021data}. As another paradigm driven by the date science, DNNs can be trained by a large amount of analytical/experimental/simulation data, corresponding to the so-called {\it supervised learning}.
Once trained, the DNNs can offer solutions for the PDEs in interpolated and even slighly extrapolated space of parameters much faster than the traditional numerical schemes. However, data are not always abundant in realistic applications or only partially accessible at best. To address this issue, physics-informed neural networks (PINNs) are proposed and trained by combining physics laws in the form of PDEs together with available data~\supercite{raissi2019physics}. In contrast to the undecorated DNNs, PINNs can be trained with data in shortage or even missing
to solve a forward problem described by known PDEs. This corresponds to an {\it unsupervised or weakly supervised learning}.
Moreover, PINNs can also deal with an inverse problem at ease, where some coefficient values in the PDEs are unclear, such as the viscosity in the Navier-Stokes equations~\supercite{karniadakis2021physics}. 
Since its invention, there have already been many innovative works to improve the accuracy and efficiency of PINNs~\supercite{yu2022gradient,wang2022and,mattey2022novel,wang2022respecting,2021Solving,xiong2022gradient}.
Concerning the discontinuous solutions of PDEs, quite a few work have been conducted with PINNs. In the seminal work of PINNs~\supercite{raissi2019physics}, a viscous term is added to smooth the shock produced by the Burgers' equation. Mao et al.~\supercite{mao2020physics} choose to distribute more sampling points in the discontinuous region identified beforehand, forming a cluster of points for a better training. Compared with uniformly or randomly sampled distribution of points, their strategy achieves a higher precision. The conservative PINNs proposed by Jagtap et al.~\supercite{jagtap2020conservative} solve the continuous and discontinuous regions separately, where in the discontinuous part a larger network and more data are selected for training. Their work demonstrates that a special division of the training area can get more accurate solutions than that of other conventional divisions. Moreover, Patel et al. propose control-volume based PINNs~\supercite{patel2022thermodynamically} and combine them with a finite volume method, when no derivative exists in the discontinuous part. They define a loss function for a single control volume and obliterate derivative operation by integrating the equation, to ensure a correct solution. 

Different from previous works, we aim to employ PINNs to solve a forward problem of compressible flows {\it without exogenous data}, corresponding to an unsupervised learning. In particular, we are interested in a typical flow problem taking place in a so-called de Laval nozzle or converging-diverging (CD) nozzle.
Its actual solution depends on the ratio between the pressures at the outlet and inlet, and in the diverging region it may branch into subsonic flow, supersonic flow, and a mixture of both with discontinuous transition where a normal shock takes place.
Since both analytical and numerical solutions for steady states exist in textbooks~\supercite{white2016, anderson1995}, 
this problem serves as an ideal benchmark to examine PINNs' performance for continuous and discontinuous compressible flows.
A thorough understanding of the procedure for seeking solutions of this simple problem may provide a guidance to solve more general compressible flows, whether there is auxiliary data or not.

\begin{figure}[htbpp]
\centering { \subfigure[]{\includegraphics[width=65mm]{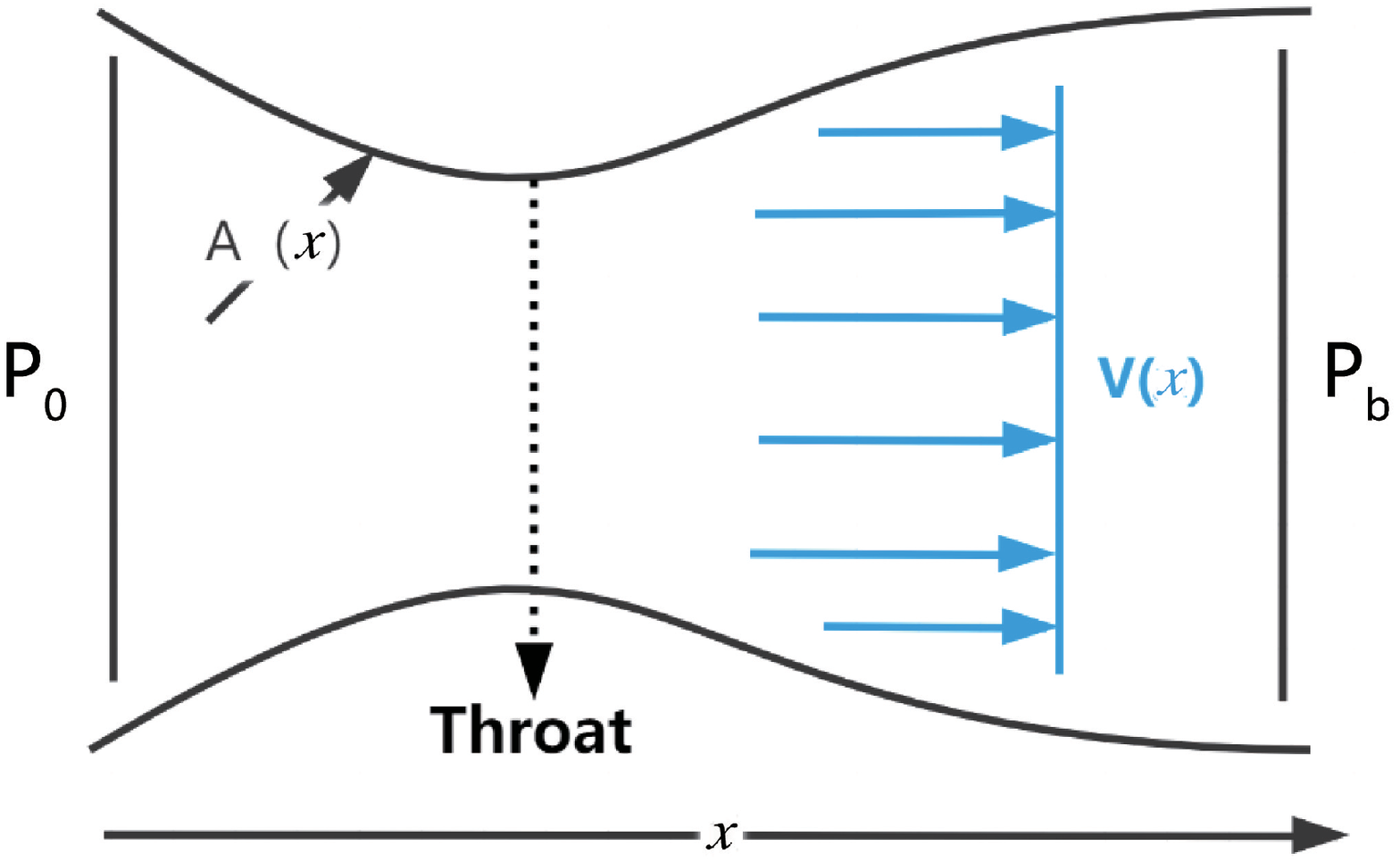}\label{fig a}}\quad 
\subfigure[]{\includegraphics[width=65mm]{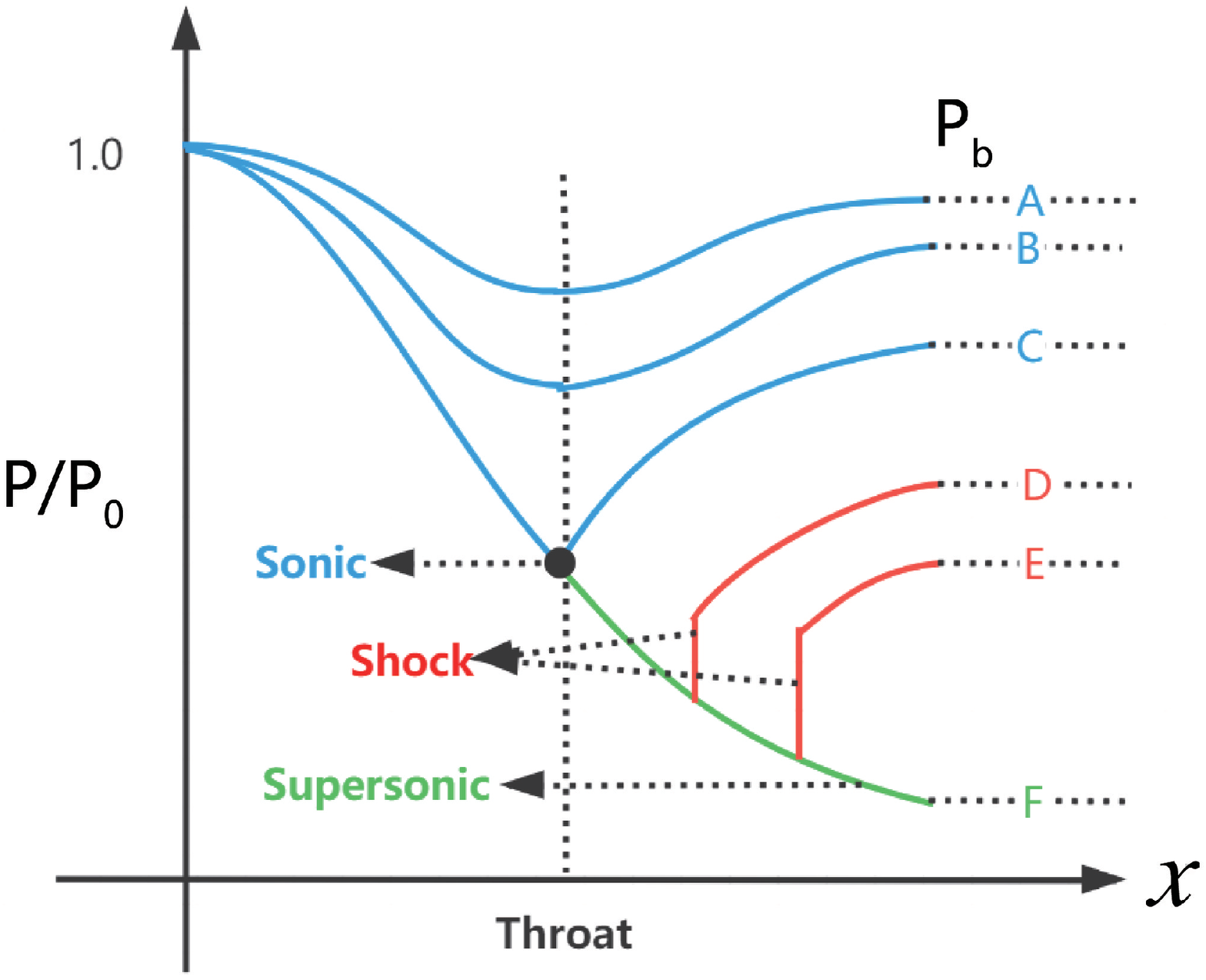}\label{fig b}}} 
\caption{ (a) one dimensional approximation to the flow within a converging-diverging (CD) nozzle; (b) pressure distributions and flow characteristics along the CD nozzle according to various ratios between the back/outlet pressure and stagnation/inlet pressure. Curves A and B correspond to subsonic flows; curve F corresponds to a subsonic-supersonic flow; curve D and E correspond to flows with norm shocks in-between supersonic-subsonic transitions.}\label{fig1}
\end{figure}
A typical three-dimensional CD nozzle with axis-symmetry is shown in Fig.~\ref{fig a}. We assume that the velocity has only one component in the x direction $V(x)$ and it changes with the area $A(x)$ of the cross section, which varies smoothly and slowly. 
Therefore, one-dimensional equations are adequate to describe the flows.
The smallest area named as throat controls the flow rate
when the flow is supersonic in the diverging part.
According to the pressure ratio between the back pressure $P_{b}$ at the outlet and stagnation pressure $P_0$ at the inlet,  flow characteristics vary, as shown in Fig.~\ref{fig b}. 
In particular, flows with a mixture of subsonic-supersonic-subsonic properties take place,  as curves D or E, when $P_b$ is moderate. Normal shock waves are generated for the transition from supersonic back to subsonic flows to raise the inner pressure up to conform with the back pressure at the outlet. The generation of shock wave is a non-isentropic process and manifests itself as a discontinuity in the context of continuum mechanics, which renders reliable solutions of the traditional numerical methods difficult.

In this work, we adopt a specific version of PINNs with a universal setting to resolve the flows in the CD nozzle, whether there is a shock wave or not, without any auxiliary data. By varying $P_b$, it offers different branches of solutions correctly. If there is a normal shock, it identifies the shock location accurately and meanwhile provides a sharp solution. 
The structure of the paper is arranged as follows. In Sec.~\ref{sec2}, flow equations together with the structure and parameters of PINNs are introduced. In Sec.~\ref{sec3}, hard constraints on the boundaries are introduced and weights of different loss functions are universally determined so that different flow characteristics are predicted at steady states. In Sec.~\ref{sec4} unsteady flows are solved and the influence of different hyper-parameters of the neural networks is analyzed to successfully improve the accuracy of solutions. In Sec.~\ref{sec5}, governing equations in conservative form are discussed and finally, in Sec.~\ref{sec6} a summary is made.

\section{The method}\label{sec2}

The one-dimensional Euler's equations are adopted to describe the flows and more specifically, a simple differential form is initially considered 
\begin{align}
 \label{eq_euler}
 \left \{\aligned
 A\frac{\partial \rho}{\partial t}+Av\frac{\partial \rho }{\partial x} +A\rho\frac{\partial v }{\partial x}+v\rho\frac{\partial A }{\partial x}=0,\\
 A\rho\frac{\partial v}{\partial t}+Av\rho\frac{\partial v }{\partial x}+A\frac{\partial P}{\partial x} =0,\\
 A\rho\frac{\partial T}{\partial t}+Av\rho\frac{\partial T }{\partial x}+P\left ( A\frac{\partial v}{\partial x}+v\frac{\partial A}{\partial x} \right )  =0 ,\\
 P-\rho RT =0.
 \endaligned
 \right.
 \end{align} 
Here $\rho$, $v$,  $T$, and $P$ are density, velocity, temperature, and pressure, respectively. 
$R$ is the universal gas constant.
We fix the stagnation properties such as density $\rho_0$, temperature $T_0$, and pressure $P_0$ at the inlet~(left) and adjust the the back pressure $P_{b}$ at the outlet~(right), to generate different flow regimes in the CD nozzle, as shown in Fig.~\ref{fig1}. A typical set of stagnation values are 
$\rho_0 = 1.52 kg/m^{3}$, $T_0 = 286.1 K$, and $P_0 = 1.247\times 10^{5} N/m^{2}$, respectively.

Flow properties are made dimensionless by $\rho_0$, $T_0$ and throat area $A^*$ as follows:
$$\rho '  = \frac{\rho }{\rho _{0} } , T'  = \frac{T}{T_{0} } , A'=\frac{A}{A^{*}}. $$
Therefore, the sound speed at stagnation is $a_{0}=\sqrt{\gamma RT_0}$, where $\gamma=1.4$ is the specific-heat ratio of air. Furthermore, $P' = P/P_{0}$,  $v'=v/a_{0}$,  and $x'=x/\sqrt{A^*}$, $t'=ta_{0}/\sqrt{A^*}$.
Eqs.~(\ref{eq_euler}) become dimensionless as
 \begin{align}
 \left \{\aligned
 A'\frac{\partial \rho'  }{\partial t' } +A' v' \frac{\partial \rho'  }{\partial x' } +A' \rho' \frac{\partial v'  }{\partial x' }+v\rho\frac{\partial A'  }{\partial x' }=0,\\
 A'\left (\gamma  \rho' \frac{\partial v'  }{\partial t' }+\gamma  v' \rho' \frac{\partial v' }{\partial x' }+\frac{\partial P' }{\partial x' }\right )  =0,\\
  A'  \rho' \frac{\partial T'  }{\partial t' }\left (\frac{1}{\gamma-1 }\right )+A' v' \rho' \frac{\partial T'  }{\partial x' }\left (\frac{1}{\gamma-1 }\right )+P' \left ( A' \frac{\partial v' }{\partial x' }+v\frac{\partial A' }{\partial x' } \right )  =0, \\
  P'-\rho' T{'} = 0.
 \endaligned
 \right.
 \label{eq_euler_dimensionless}
 \end{align}
In the rest of this paper, all physical quantities shall appear in dimensionless form.
Therefore, we remove all `` $'$ "s in the equations for convenience. 

\begin{figure}
    \centering
    \includegraphics[width=0.7\columnwidth]{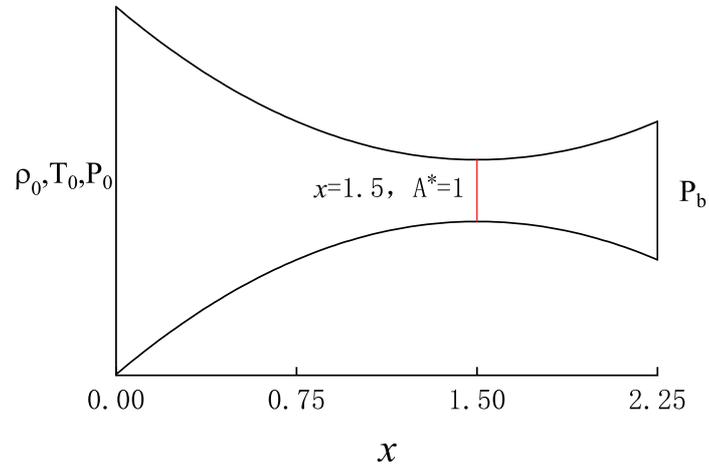}
    \caption{A converging-diverging nozzle with cross-section area $A$ described by a parabolic function of $x$ along the axis with $x\in(0,2.25)$, where the throat has the minimum area $A^*=1$ at $x=1.5$.
    $\rho _{0}$, $T_{0}$ and  $P_{0}$ are the stagnation density, temperature, and pressure at the inlet~(left), respectively. $P_{b}$ is the back pressure at the outlet~(right). We adjust $P_{b}$ to vary the pressure ratio and to generate different flow regimes in the nozzle. }
    \label{fig_geometry}
\end{figure}
We define the cross-section of the CD nozzle as a parabolic function along the $x$ axis: $A(x)=1+2.2*(x-1.5)^{2}$, where the minimum area $A^*=1$ takes place at $x=1.5$, as shown in Fig.~\ref{fig_geometry}. This geometry is nothing special and only taken for convenience. The methodology presented later also applies to more general geometries.

\begin{figure}
\centering
\includegraphics[width=100mm]{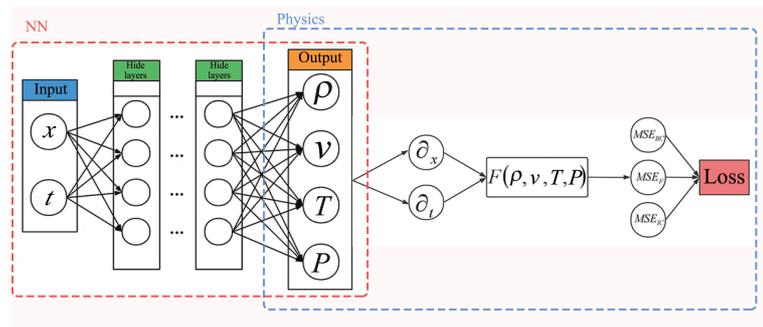}
\caption{The structure of PINNs. On the left is a simple feedforward neural network to be trained while on the right is the physics information expressed in PDEs. A loss function composed of boundary conditions, initial conditions, and physics equations together guides the training of the neural network.}
 \label{fig3}
\end{figure}
The structure of the neural networks~(NNs) is presented in Fig.~\ref{fig3},
where it is trained without any auxiliary data, except for the boundary and initial conditions. The loss function of the NNs is expressed as follows
\begin{align}
Loss=Loss_{BC}+Loss_{IC}+Loss_{F},
\end{align}
where $Loss_{BC}$, $Loss_{IC}$, and $Loss_{F}$ correspond to the sub-loss function of boundary conditions,  initial conditions, and PDEs, respectively. 
The sub-loss function of the PDEs has four components originated from the four equations in Eqs.~(\ref{eq_euler_dimensionless})
\begin{align}
 \left \{\aligned
 A\frac{\partial \rho  }{\partial t } +A v \frac{\partial \rho  }{\partial x } +A \rho \frac{\partial v  }{\partial x }+v\rho\frac{\partial A  }{\partial x }=F_1(x,t),\\
 A\left (\gamma  \rho \frac{\partial v }{\partial t }+\gamma  v \rho \frac{\partial v  }{\partial x }+\frac{\partial P }{\partial x }\right )  =F_2(x,t),\\
  A  \rho \frac{\partial T  }{\partial t }\left (\frac{1}{\gamma-1 }\right )+A v \rho \frac{\partial T  }{\partial x }\left (\frac{1}{\gamma-1 }\right )+P \left ( A \frac{\partial v }{\partial x }+v\frac{\partial A }{\partial x } \right )  =F_3(x,t), \\
  P-\rho T = F_4(x,t).
 \endaligned
 \right.
 \label{eq_loss}
\end{align}
Here $F_1(x,t)$, $F_2(x,t)$, $F_3(x,t)$, and  $F_4(x,t)$ represent residuals of the mass, momentum, energy, and state equations, respectively. 
We assume that pressure $P$ and density $\rho$ are two individual variables so that
the NNs have both variables as outputs.
Accordingly, $P$ and $\rho$ form a sub-loss function $F_4(x, t)$ via the residual from the equation of state.
Each component of the loss function has an associated weight $\omega_i$ and in part they determine the optimization of the network parameters:
\begin{align}
  F=\omega _1 F_1+ \omega _2 F_2+ \omega _3 F_3+ \omega _4 F_4.
\end{align}
By default $w_i=1$ for $i=1,2,3,4$.
All the sub-loss functions are expressed in the form of mean squared errors (MSEs)
\begin{align}
Loss_{BC} &= MSE_{BC},\\
Loss_{IC} &= MSE_{IC},\\
Loss_{F} &= MSE_{F} = \sum_{i=1}^{4} \frac{\omega_i}{N_{F} } \sum_{j = 1}^{N_{F}} \left | F_{i}\left ( x_{j} ,t_{j}  \right )\right |^{2},
\end{align}
where $F_i\left ( x_{j} ,t_{j}  \right )$ is the corresponding residual at point $(x_j, t_j)$ among all $N_F$ training poings.
The actual expressions of $Loss_{BC}$ and $Loss_{IC}$ are problem-dependent and shall be described in later sections. Because the NNs adopt the chain rule of derivatives, we do not need to approximate the partial differential terms by any special numerical scheme as in the traditional numerical methods. In the process of minimizing the loss function towards zero, the back propagation algorithm optimizes the parameters (weights and biases) of the NNs.
In this context, the optimization process is also named as training.
Once trained, the NNs can predict values $\rho$, $v$, $T$, and $P$ for any given $x$ and $t$.

\section{Steady state solutions} \label{sec3}
We commence to solve for flow problems within the CD nozzle at steady states, therefore the terms containing partial derivative with respect to time in Eqs.~(\ref{eq_euler_dimensionless}) and (\ref{eq_loss}) are temporarily discarded. For the steady states, we can obtain accurate solutions via analytical methods and make use of them to evaluate the performance of PINNs. In Sec.~\ref{sec3.1} we solve the flows in the diverging part of the nozzle by imposing the critical states of the throat as boundary conditions. In Sec.~\ref{sec3.2}, we capture the shock wave by modifying the NNs and obtain accurate solutions at a high resolution of sampling points. In Sec.~\ref{sec3.3}, we investigate the effects due to the number of training points on the accuracy of solutions.

 \subsection{Diverging channel}\label{sec3.1}

The flow characteristics are relatively simple in the converging part of the CD nozzle,
whereas it is rather complex in the diverging part.
Therefore, we first impose the physical quantities at the throat as inlet boundary conditions and calculate flows in the diverging part alone.
The analytical solutions to this problem are expressed as follows,
 \begin{align}
 \left \{\aligned
  \left [ {1+\frac{1}{2} \left ( \gamma -1 \right )Ma^{2}  } \right ]^{\frac{1 }{\left ( \gamma -1 \right )   }  }= \frac{1}{\rho },\\
  \frac{1}{Ma} \left [ \frac{1+\frac{1}{2} \left ( \gamma -1 \right )Ma^{2}  }{\frac{1}{2} \left ( \gamma +1 \right ) }  \right ]^{\frac{1}{2}\left ( \gamma +1 \right )\left ( \gamma -1 \right )   }=\frac{1}{A^* },\\
{1+\frac{1}{2} \left ( \gamma -1 \right )Ma^{2}  }=\frac{1}{T },\\
\left [ {1+\frac{1}{2} \left ( \gamma -1 \right )Ma^{2}  } \right ]^{\frac{\gamma }{\left ( \gamma -1 \right )   }  }=\frac{1}{P },
\endaligned
 \right.
  \label{eq_continuous}
 \end{align}
which are valid for both subsonic and supersonic flows.
As they are not in explicit form, some iterative procedures are necessary and
we adopt a web-based applet to calculate accurate solutions with $8$ decimal digits as references~\supercite{calculate}.

More specifically, we take the critical state of air reaching the speed of sound at the throat. Accordingly, the inlet boundary conditions for the diverging channel are
    $$(\rho _{x=1.5},v_{x=1.5},T_{x=1.5} ,P_{x=1.5})=(0.634, 0.912, 0.833, 0.528).$$
Given the geometry, analytical solution may be subsonic, supersonic and a mixture of both with a discontinuous shock.
Both subsonic and supersonic solutions are smooth and unique, such as $C$ and $F$ curves on Fig.~\ref{fig b}, and are available from Eqs.~(\ref{eq_continuous}).
Moreover, there are infinitely many solutions, each of which has a discontinuous shock,
such as $D$ and $E$ curves on Fig.~\ref{fig b} and more are not shown.
Each solution is unique corresponding to one specific boundary condition at the outlet.
Out of curiosity, however, we intentionally leave the outlet boundary for free,
to interrogate what PINNs generate.

The computational domain is for $x\in \left [ 1.5,2.25 \right ] $ and the loss function expressed as MSEs for the inlet boundary is defined as 
 \begin{align}
    MSE_{BC} =MSE_{\rho}+MSE_{v}+MSE_{T}+MSE_{P}, 
 \end{align}
 with:
 \begin{align}
 \gathered
MSE_{\rho} = \frac{1}{N_{BC} } \sum_{j = 1}^{N_{BC}} (\left | \rho_{NN}\left ( x_{j}=1.5  \right )-\rho(x_{j}=1.5)\right |^{2}),\\
 MSE_{v} = \frac{1}{N_{BC} } \sum_{j = 1}^{N_{BC}} (\left | v_{NN}\left ( x_{j}=1.5  \right )-v(x_{j}=1.5)\right |^{2}),\\
 MSE_{T} = \frac{1}{N_{BC} } \sum_{j = 1}^{N_{BC}} (\left | T_{NN}\left ( x_{j}=1.5  \right )-T(x_{j}=1.5)\right |^{2}),\\
 MSE_{P} = \frac{1}{N_{BC} } \sum_{j = 1}^{N_{BC}} (\left | P_{NN}\left ( x_{j}=1.5  \right )-P(x_{j}=1.5)\right |^{2}).
 \endgathered
 \end{align}
Here the values with subscript ``$_{NN}$" are the ones predicted by the NN, while the bare values are imposed ones. NNs' parameters for weight and bias are initialized by Glorot scheme~\supercite{glorot2010understanding}. This convention also applies later on. 
The NN has 3 hidden layers, each layer has 30 neurons, and tanh is the activation function.
We choose $N_{BC}=1$ and $N_F=100$ points in the x direction,
for the inlet boundary and collocation points within the physical domain, respectively.
To investigate clearly the influence of other parameters on the results, we intentionally choose all training points uniformly distributed.

\begin{figure}[htbp]
    \centering
    \includegraphics[width=0.8\columnwidth]{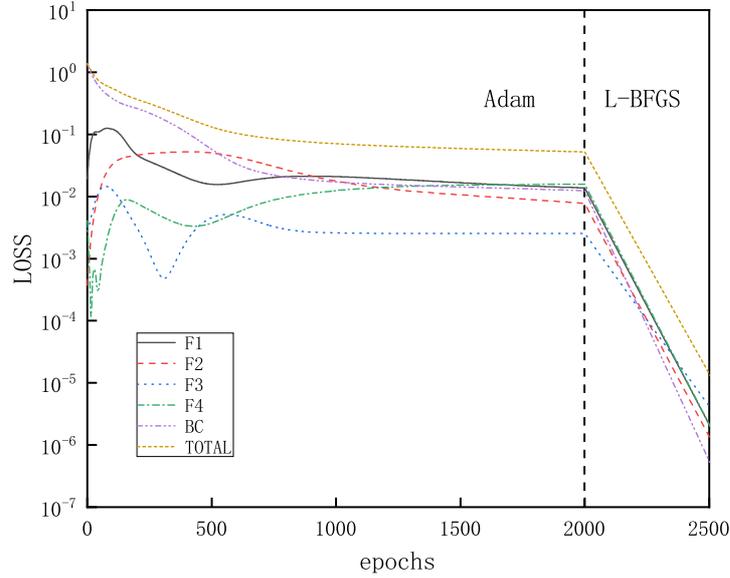}
    \caption{Evolution of the loss function for a typical instance of PINNs for solving the flows in the diverging channel: with Adam and L-BFGS optimizers, each component of the loss function descends towards minimization with more training epochs.}
    \label{fig_l0}
\end{figure}
The training proceeds with Adam optimizer of learning rate $0.0001$ for $2000$ epochs and
continues with L-BFGS optimizer for $500$ epochs~\supercite{kingma2014adam,zhu1997algorithm}.  
A typical evolution of the loss function is shown in Fig.~\ref{fig_l0},
where each component descends clearly towards minimization with more training epochs.
After completion of the training, $50$ evenly distributed points in the computational domain are selected for predicting  $\rho$, $Ma=v/c=v/\sqrt{\gamma R T}$, $T$ and $P$,
which corresponds to a simple forward pass of the NNs.
Since both sets of points are evenly distributed, the prediction points are actually a subset of the training points. 
This convention also applies later on, unless otherwise stated.

\begin{figure}[htbp]
\centering {  \subfigure[Density]{\includegraphics[width=0.48\columnwidth]{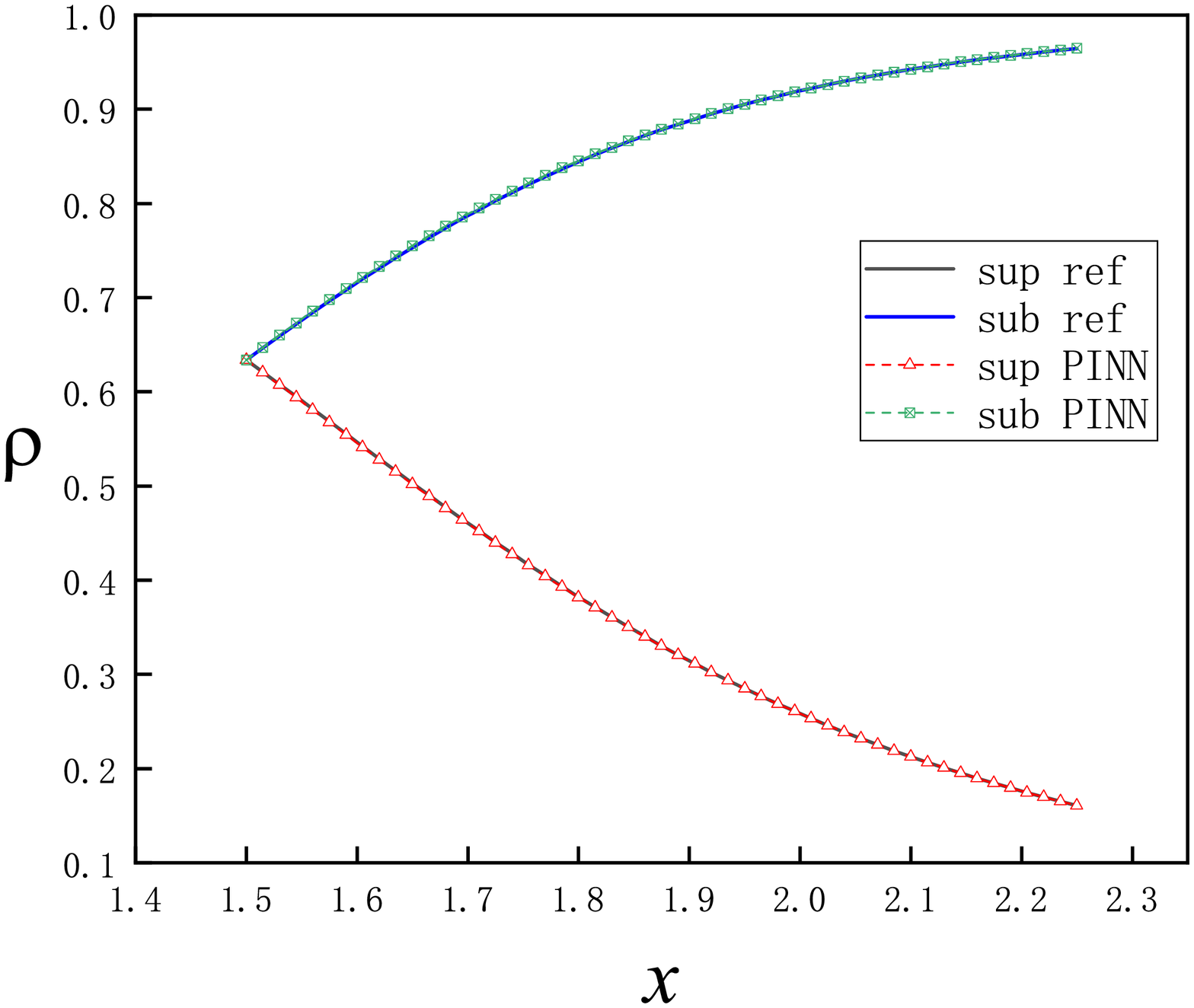}}\quad
\subfigure[Mach number]{\includegraphics[width=0.48\columnwidth]{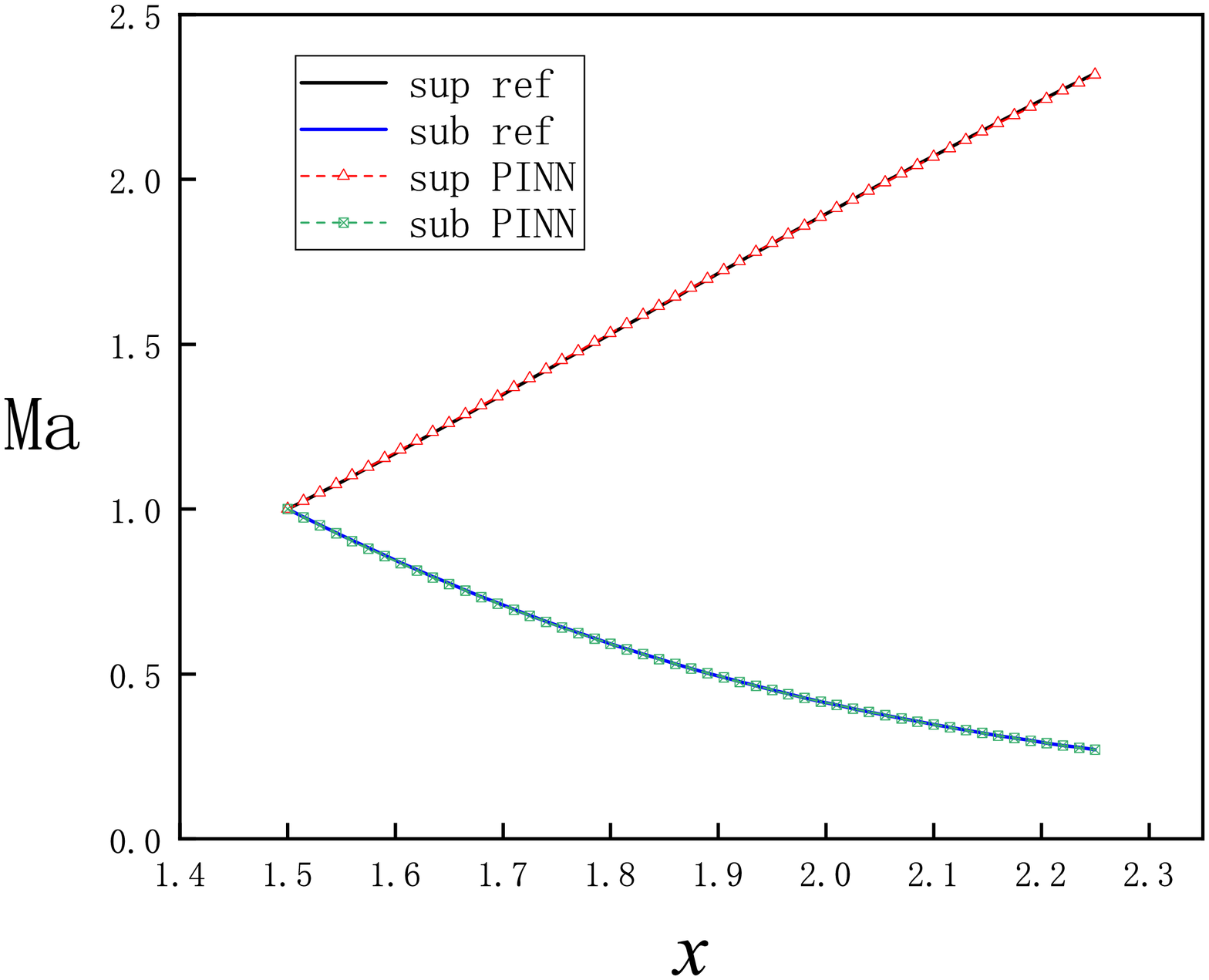}}
\subfigure[Temperature]{\includegraphics[width=0.48\columnwidth]{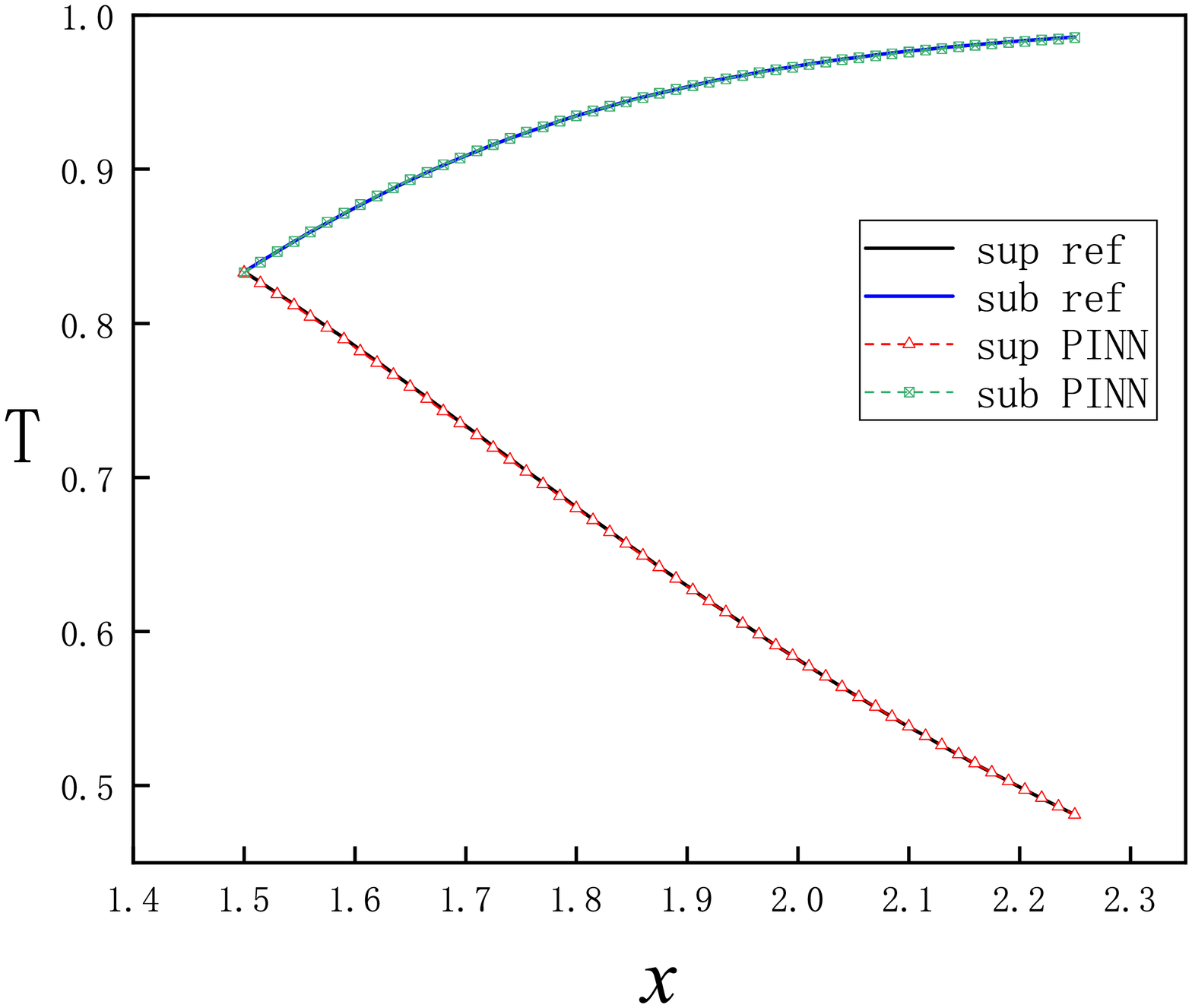}}\subfigure[Pressure]{\includegraphics[width=0.48\columnwidth]{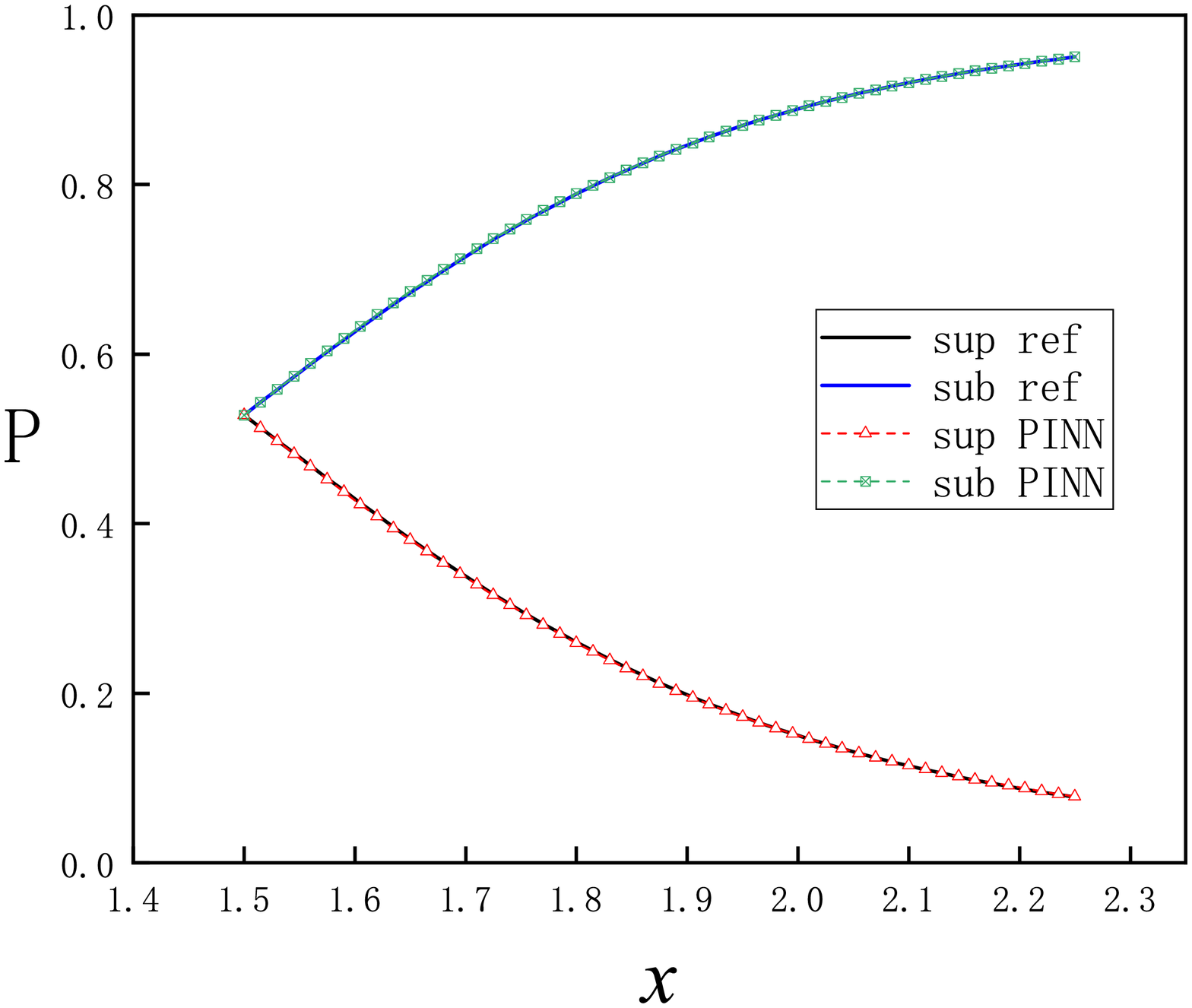}}} \caption{Solutions for flows in the diverging channel. Only inlet boundary conditions are imposed while the outlet boundary conditions are intentionally left free. Each running instance of PINNs generates an accurate supersonic or subsonic solution randomly, due to the inherent randomness during the initialization of the NN's parameters.
Furthermore, PINNs deliberately avoid the subtle branch of infinite many solutions, which involve discontinuous shocks.}\label{fig_suiji}
\end{figure}
With the same setting,  we run multiple instances of PINNs to predict the flow solutions.
Surprisingly, for each instance we may obtain one of two different solutions randomly.
One is for supersonic flows and the other is for subsonic flows,
as shown in Fig.~\ref{fig_suiji}.
Moreover, both the subsonic and supersonic solutions of PINNs are in excellent agreement with the analytical ones, described by Eqs.~(\ref{eq_continuous}).
After many instances of training and prediction, this observation is repeatable,
that is, without providing the outlet boundary condition, 
PINNs are always able to find one of the two smooth solutions.
We attribute this uncertain outcome to the random initialization of the NN's parameters~\supercite{glorot2010understanding}. 
Nevertheless, without an appropriate boundary condition for the outlet, PINNs deliberately circumvent the discontinuous solutions.

\subsection{Converging-diverging nozzle}\label{sec3.2}
We continue to consider the whole geometry of the CD nozzle.
Stagnation values of density, temperature, and pressure are given at the inlet,
while a back pressure $P_b$ is provided at the outlet.
Depending on the value of $P_b$, different flow characteristics occur.
For smooth solutions of both subsonic and supersonic flows, 
the analytical expressions in Eq.~(\ref{eq_continuous}) are utilized.
For a solution with a normal shock in the diverging part, we refer to the
Rankine-Hugoniot equations as follows:
 \begin{align}
 \left \{\aligned
 \frac{(\gamma+1)Ma_{1}^{2} }{(\gamma-1)Ma_{1}^{2} +2} = \frac{\rho_1}{\rho_2},\\
  \frac{(\gamma-1)Ma_{1}^{2} +2}{2\gamma Ma_{1}^{2}-(\gamma-1)} = 
 Ma_{2}^{2} ,\\
 [2+(\gamma - 1)Ma_{1}^{2} ]\frac{2 \gamma Ma_{1}^{2}-(\gamma -1)}{(\gamma +1)^2Ma_{1}^{2}} = \frac{T_1}{T_2},\\
 \frac{1}{\gamma +1} [2 \gamma Ma_{1}^{2}-(\gamma-1) ] = \frac{P_1}{P_2},
\endaligned
 \right.
  \label{eq_rh}
 \end{align}
which relates the physical values before and after the shock.
A similar solution strategy is adopted to calculate two accurate and smooth solutions
with $8$ decimal digits pieced together at the shock as references~\supercite{calculate}.

The computational domain is: $x\in \left [ 0,2.25 \right ]$.
The following boundary conditions are applied in PINNs:
 $$(\rho _{x=0},T_{x=0} ,P_{x=0})=(1.0, 1.0, 1.0), \quad  P_{x=2.25}=P_{b}.$$
Therefore, the loss function for the boundary conditions is expressed as MSEs as follows:
 \begin{align}
     MSE_{BC} =MSE_{\rho}+MSE_{T}+MSE_{P}
 \end{align}
 with:
 \begin{align}
 \gathered
MSE_{\rho} = \frac{1}{N_{BC} } \sum_{j = 1}^{N_{BC}} (\left | \rho_{NN}\left ( x_{j}=0  \right )-\rho(x_{j}=0)\right |^{2}),\\
 MSE_{T} = \frac{1}{N_{BC} } \sum_{j = 1}^{N_{BC}} (\left | T_{NN}\left ( x_{j}=0  \right )-T(x_{j}=0)\right |^{2}),\\
 MSE_{P} = \frac{1}{N_{BC} } \sum_{j = 1}^{N_{BC}} (\left | P_{NN}\left ( x_{j}=0  \right )-P(x_{j}=0)\right |^{2}+\left | P_{NN}\left ( x_{j}=2.25  \right )-P(x_{j}=2.25)\right |^{2}).
 \endgathered
 \end{align}
The NN has 3 hidden layers and each layer 30 neurons.
$N_{BC}=2$ and $N_F=200$ are evenly distributed in the $x$ direction.
The training proceeds with Adam optimizer for $2000$ epochs and continues with L-BFGS optimizer for $500$ epochs.  
After training, $50$ evenly distributed points are selected for predicting solutions.

\begin{figure}[htbp]
\centering { \subfigure[Density]{\includegraphics[width=0.48\columnwidth]{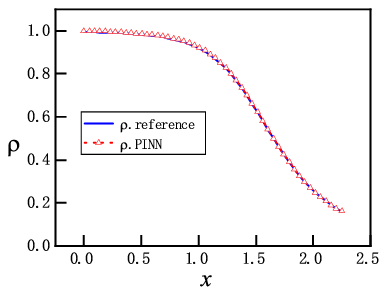}}\quad
\subfigure[Mach number]{\includegraphics[width=0.48\columnwidth]{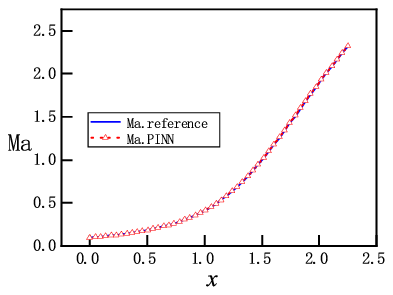}}
\subfigure[Temperature]{\includegraphics[width=0.48\columnwidth]{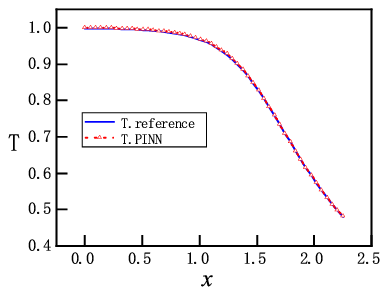}}
\subfigure[Pressure]{\includegraphics[width=0.48\columnwidth]{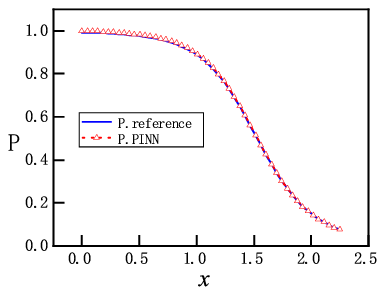}}\quad 
} \caption{PINNs' solutions for supersonic continuous flows: with $P_b=0.07726$, the solutions of $\rho$, $Ma$, $T$, and $P$ are smooth, with subsonic flow in the converging part while supersonic flow in the diverging part. References are taken from the analytical solutions governed by Eq.~(\ref{eq_continuous}).}\label{fig_supersonic}
\end{figure}

Firstly, $P_b=0.07726$, the flow is subsonic in the converging part and completely supersonic in the diverging part.
The solutions from PINNs are shown in the Fig.~\ref{fig_supersonic},
where the reference solutions according to Eqs.~(\ref{eq_continuous}) are also presented for comparison.
We observe that PINNs with a default setting offer excellent accuracy for this type of compressible flow, where there is smooth transition from a subsonic flow to a supersonic flow at the throat $x=1.5$,
corresponding to curve F on Fig.~\ref{fig b}.

\begin{figure}[htbp]
\centering { \subfigure[Density]{\includegraphics[width=0.48\columnwidth]{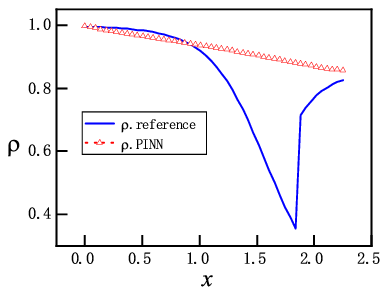}}\quad
\subfigure[Mach number]{\includegraphics[width=0.48\columnwidth]{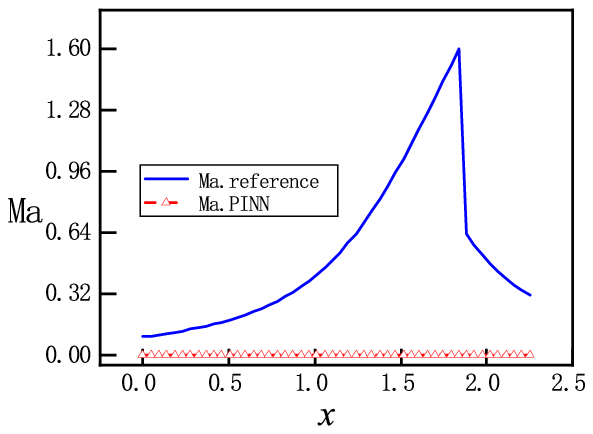}}\quad
\subfigure[Temperature]{\includegraphics[width=0.48\columnwidth]{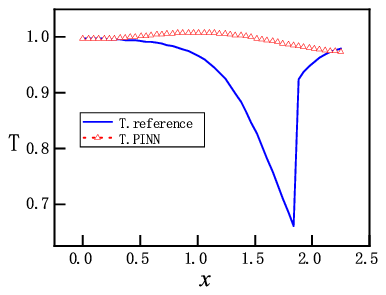}}\quad
\subfigure[Pressure]{\includegraphics[width=0.48\columnwidth]{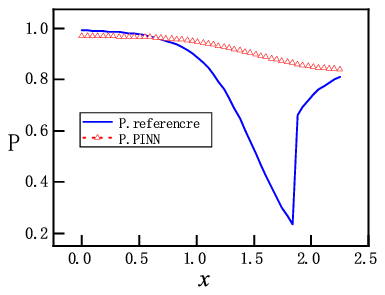}}} \caption{PINNs' solutions for subsonic and supersonic discontinuous flows. With $P_b=0.81017$, the analytical solutions of $\rho$, $Ma$, $T$, and $P$ are governed by by Eqs.~(\ref{eq_continuous}) and (\ref{eq_rh}). Solutions should be smooth and subsonic in the converging part and become discontinuous at $x=1.875$ in the diverging part,
where a normal shock wave is expected. PINNs with default settings fail to reproduce the correct solutions and increasing the number of training points and/or epochs does not help.}\label{fig_wrong}
\end{figure}
\begin{figure}[htbp]
    \centering
    \includegraphics[width=0.8\columnwidth]{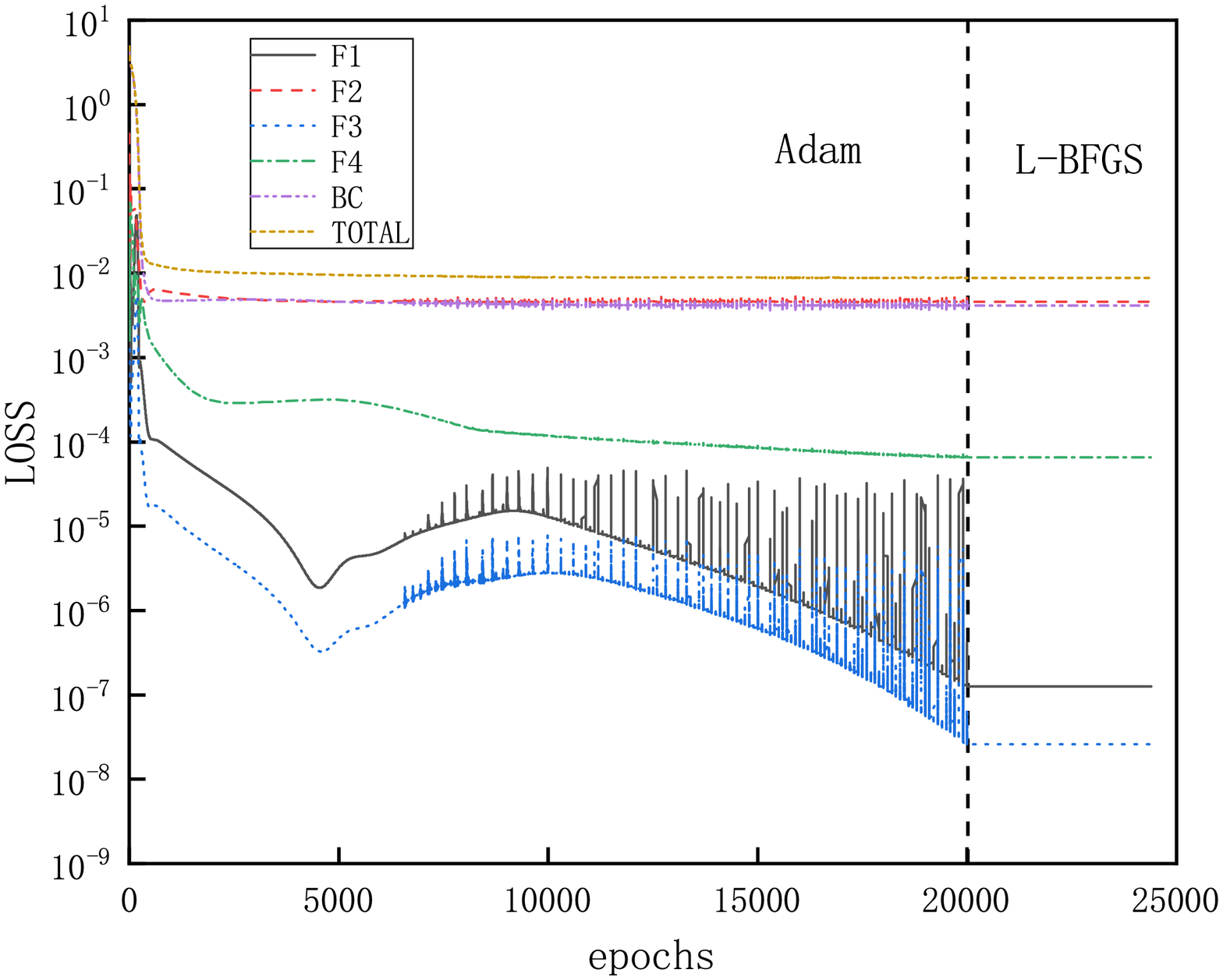}
    \caption{Evolution of the loss function of PINNs for solving the flows in the CD nozzle with $P_b=0.81017$, where a normal shock is expected in the diverging part: with Adam and L-BFGS optimizers, components of the loss function are reluctant to descend towards minimization, especially those for boundary (BC) and momentum equation (F2).}
    \label{fig_l1}
\end{figure}
Furthermore, we set $P_{b}=0.81017$. According to Eqs.~(\ref{eq_continuous}) and (\ref{eq_rh}), a normal shock wave is expected at $x=1.875$ in the diverging part of the nozzle.
With a default setting, PINNs offer solutions that are far away from the references, as shown in Fig.~\ref{fig_wrong}.
An examination at the evolution of the loss function during training
reveals that the troublemakers are the losses for the boundary condition and momentum equation, as shown in Fig.~\ref{fig_l1}.
Increasing the number of training epochs for ten times from $2500$ to $25000$ epochs does not improve the descent of the loss function.
Enhancing the number of sampling points ten times from $200$ to $2000$ does not help either (results are not shown).

A few notes are in order.
Since we employ PINNs as a direct numerical simulation tool without prior data,
they not aware of the shock location
in advance and can not distribute more sampling points around the shock.
This renders an accurate prediction of the discontinuous flow by PINNs challenging. 
After a closer inspection on Fig.~\ref{fig_wrong}(b),
we observe that the velocity/Mach number profile is almost flat around zero.
This indicates that during the training the optimizer of NNs 
ignores the residual of momentum equation.
Leaving a flat velocity implies all velocity terms can be
discarded from $F_1$, $F_2$ and $F_3$ in Eq.~(\ref{eq_loss}),
since $\partial v/\partial x \approx 0$.
We interpret $v(x)=constant$ as a trivial, but wrong solution.
For the loss of $F_1$ and $F_3$, the term $\partial A/\partial x$ 
is known from the geometry, which demands the residuals towards zero,
as shown in Fig.~\ref{fig_l1}.
However, with the trivial solution of $v(x)=constant$,
the pressure cannot satisfy two Dirichlet boundary conditions
for the inlet and outlet simultaneously 
that is , $\partial P/\partial x=0$ is impossible, as shown in Fig.~\ref{fig_wrong}(d).
With $\partial v/\partial x \approx 0$, $\partial P/\partial x$ is the only term left in $F_2$,
which does not descend towards zero.
Consequently, the losses for boundary condition and $F_2$ are both large
and do not descend easily, as shown in Fig.~\ref{fig_l1}.
\begin{figure}[htbp]
    \centering
    \includegraphics[width=0.8\columnwidth]{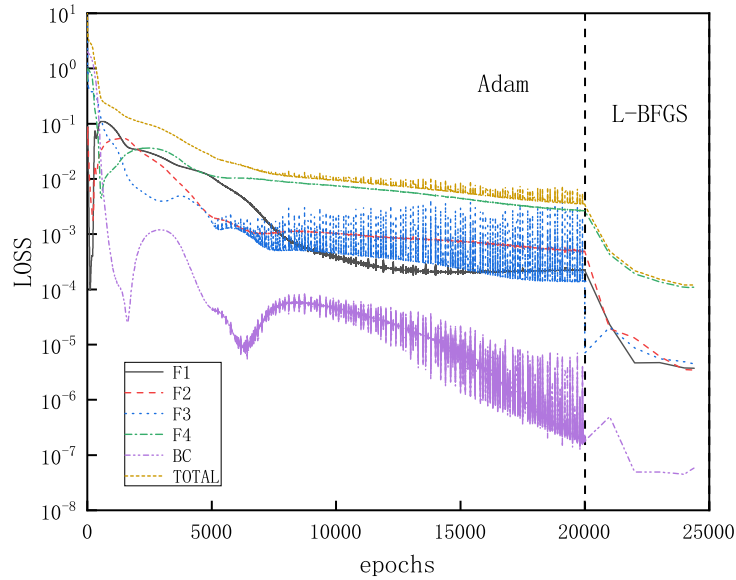}
    \caption{Evolution of the loss function of PINNs for solving the flows in the CD nozzle with $P_b=0.81$, where a normal shock is expected in the diverging part: with Adam and L-BFGS optimizers. With more weight on the momentum equation and hard constrain on the Dirichlet boundary conditions, all components of the loss function descend easily.}
    \label{fig_l2}
\end{figure}
 \begin{figure}[htbp]
\centering { \subfigure[Density]{\includegraphics[width=0.48\columnwidth]{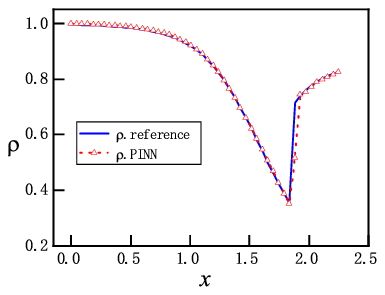}}\quad
\subfigure[Mach number]{\includegraphics[width=0.48\columnwidth]{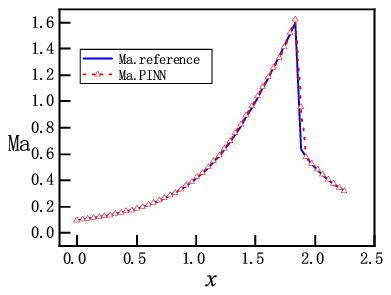}}\quad
\subfigure[Temperature]{\includegraphics[width=0.48\columnwidth]{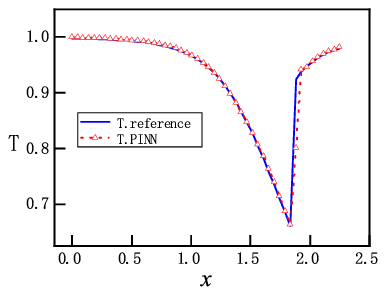}}\quad
\subfigure[Pressure]{\includegraphics[width=0.48\columnwidth]{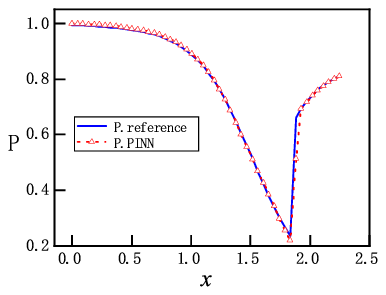}} } \caption{PINNs' solutions for subsonic and supersonic flows with discontinuity. With $P_b=0.81017$, the analytical solutions of $\rho$, $Ma$, $T$, and $P$ are governed by by Eqs.~(\ref{eq_continuous}) and (\ref{eq_rh}). Solutions should be smooth and subsonic in the converging part and become discontinuous at $x=1.875$ in the diverging part,
where a normal shock wave is expected. PINNs with a proper weight on the momentum loss function and hard constraint on the Dirichlet boundary conditions identify shock location and reproduce accurately discontinuous flows at ease.}\label{fig_shock}
\end{figure}

Motivated by the observations and speculations above,
we introduce two small modifications into the vanilla version PINNs.
Firstly, we choose to redistribute the weights between the components of the loss function
for the PDEs as $\omega_{1}:\omega_{2}:\omega_3:\omega_4=1:20:1:1$,
to enhance the descent of momentum residual.
In addition, we employ hard constraints of pressure to satisfy the Dirichlet boundary conditions exactly,
instead of minimizing the MSEs. 
The NN still has $3$ hidden layers and each layer has $30$ neurons. 
Moreover, $2000$ uniformly distributed points in the x direction are selected as training points.
The training proceeds with Adam for $20000$ epochs and continues with L-BFGS for $5000$ epochs.
The upgraded evolution of loss function during training is shown in Fig.~\ref{fig_l2},
where a clear descent for all components is observed.
Furthermore, new predictions of PINNs are in Fig.~\ref{fig_shock},
where a normal shock wave is observed at $x=1.875$ and meanwhile
sharp profiles are reproduced accurately for $\rho$, $Ma$, $T$ and $P$ with $50$ points.
We emphasize that with this new setting, the location of the shock is identified by PINNs automatically without any other efforts.
We note that both the modified weights and hard constraints on the Dirichlet boundary conditions
are necessary for the accurate solutions.

\begin{figure}[htbp]
\centering { \subfigure[Density]{\includegraphics[width=0.48\columnwidth]{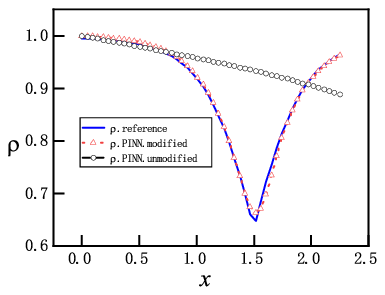}}\quad 
\subfigure[Mach number]{\includegraphics[width=0.48\columnwidth]{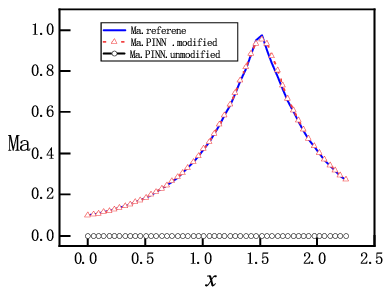}}
\subfigure[Temperature]{\includegraphics[width=0.48\columnwidth]{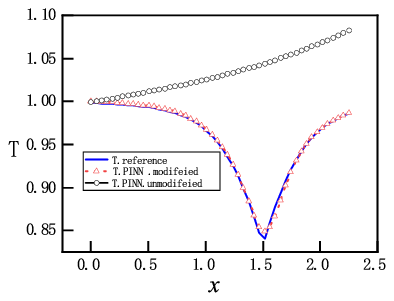}}\subfigure[Pressure]{\includegraphics[width=0.48\columnwidth]{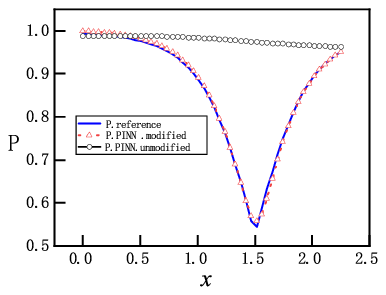}}} \caption{PINNs' solutions for subsonic flows. For $P_b=0.95055$, a subsonic flow is expected in the entire nozzle as curve B on Fig.~\ref{fig b}. PINNs with a default setting cannot deal with continuous problem properly, whereas PINNs with modified weights and hard constrains on the Dirichlet boundary conditions predict the flows accurately.}\label{fig_subsonic}
\end{figure}
Lastly, we set $P_b=0.95055$ so that a subsonic flow is expected in the entire nozzle.
Results of two versions of PINNs are presented in Fig.~\ref{fig_subsonic}.
We observe that PINNs with a default setting provides a trivial and wrong solution of $v(x)\approx 0$
and also incorrect solutions for $\rho$, $T$ and $P$.
With modified weights of on the momentum loss and hard constrains on the Dirichlet boundary conditions, 
PINNs are able to predict accurately the continuous subsonic flow in the entire nozzle,
which turns trend after the throat at $x=1.5$.
This corresponds to curve B on Fig.~\ref{fig b}.

With this new setting, PINNs are also able to reproduce Figs.~\ref{fig_suiji} and \ref{fig_supersonic},
results of which are omitted.
Therefore, we shall continue to use this version of PINNs for the rest of the work.

\subsection{Effects of resolution}\label{sec3.3}
\begin{figure}[htbp]
\centering { \subfigure[Density]{\includegraphics[width=0.48\columnwidth]{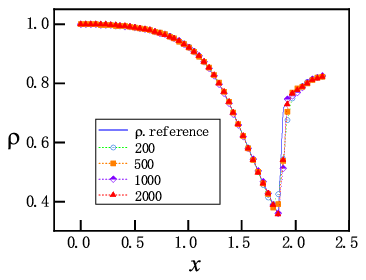}}\quad 
\subfigure[Mach number]{\includegraphics[width=0.48\columnwidth]{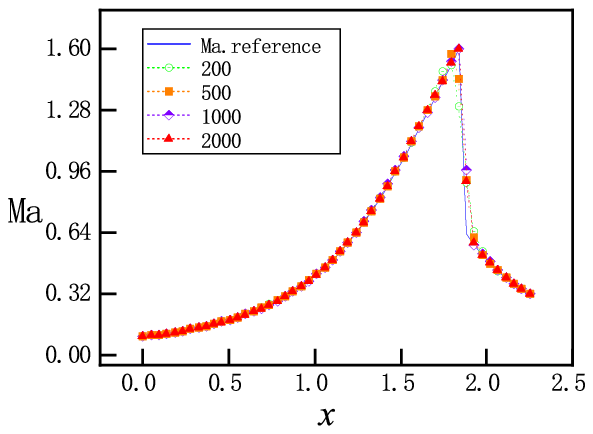}}
\subfigure[Temperature]{\includegraphics[width=0.48\columnwidth]{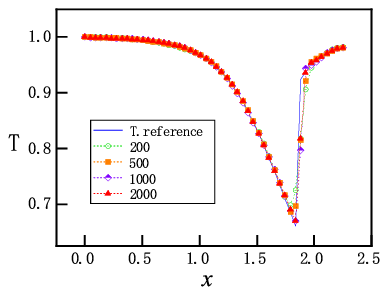}}\subfigure[Pressure]{\includegraphics[width=0.48\columnwidth]{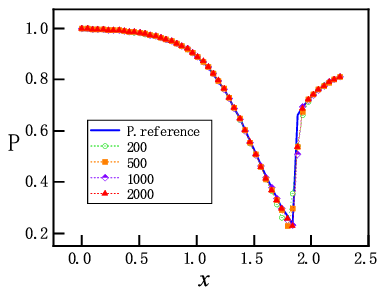}}} \caption{The number of training points is 200,500,1000 and 2000, respectively. The prediction has 50 points. They are all evenly distributed in the $x$ direction.}\label{fig_super_with_time}
\end{figure}

In this section, the effects of the number of training points for PINNs  are explored. 
We exemplify this study by $P_b=0.81017$, where a normal shock is expected in the diverging part of the nozzle.
Four different numbers of training points are selected: $200$, $500$, $1000$ and $2000$, 
which are uniformly distributed in the x direction. 
After training, physical quantities at $50$ uniformly distributed points are selected for predictions. 
As shown in Fig.~\ref{fig_super_with_time}, with all four resolutions of training,
results are stable and accurate for density, Mach number, temperature and pressure.
It is worth noting that there is no Gibbs phenomenon,
which is typically observed in traditional numerical methods.

\section{Time-dependent solutions}\label{sec4}

A more realistic flow in a CD nozzle has transit states from rest to the steady state. 
There are no analytical solutions for this time-dependent flow, 
which makes a numerical procedure necessary. 
In this section, we leverage the power of PINNs to tackle this time-dependent problem,
possibly with normal shocks.
In Sec.~\ref{sec4.1}, PINNs are employed to solve the time-dependent flows,
for which steady state is supersonic in the diverging part of the nozzle. 
In Sec.~\ref{sec4.2}, the influences of NNs' parameters, such as the size, the number of training points and different distributions of training points on the solution accuracy are discussed. 
In Sec.~\ref{sec4.3}, different initial conditions for the time-dependent flows are investigated,
for which steady state has a normal shock.

\subsection{Subsonic-supersonic continuous flow}\label{sec4.1}

In this section, we solve the unsteady flow in the nozzle
for the initial and boundary conditions given as follows

$$(\rho^{t=0}_{x},v_{x}^{t=0},T_{x}^{t=0},P_{x}^{t=0})=(1.0, 0.0, 1.0, 1.0),$$
$$(\rho^{t} _{x=0},T_{x=0}^{t} ,P_{x=0}^{t})=(1.0, 1.0, 1.0),\quad
 P_{x=2.25}^{t}=P_{b}=0.07726.$$
As shown in the previous section, the flow is supersonic in the diverging part of the nozzle for a steady flow.
For a time-dependent flow, the computational domain in space and time is $x\in \left [ 0,2.25 \right ]$ and $t\in \left [ 0,8 \right ] $, respectively. 
Furthermore, the loss function for ICs expressed as MSEs is defined as
 \begin{align}
\aligned
MSE_{IC} =MSE_{\rho}^{IC}+MSE_{v}^{IC}+MSE_{T}^{IC}+MSE_{P}^{IC}
\endaligned
\end{align}
 with:
  \begin{align}
\aligned
  MSE_{\rho}^{IC} = \frac{1}{N_{IC} } \sum_{j = 1}^{N_{IC}} (\left | \rho_{NN}\left ( x_{j},0  \right )-\rho(x_{j},0)\right |^{2}),\\
 MSE_{v}^{IC} = \frac{1}{N_{IC} } \sum_{j = 1}^{N_{IC}} (\left | v_{NN}\left ( x_{j},0  \right )-v(x_{j},0)\right |^{2}),\\ 
 MSE_{T}^{IC} = \frac{1}{N_{IC} } \sum_{j = 1}^{N_{IC}} (\left | T_{NN}\left ( x_{j},0  \right )-T(x_{j},0)\right |^{2}),\\ MSE_{P}^{IC} = \frac{1}{N_{IC} } \sum_{j = 1}^{N_{IC}} (\left | P_{NN}\left ( x_{j},0  \right )-P(x_{j},0)\right |^{2}).
\endaligned
\end{align}
In addition, the boundary conditions for pressure are implemented as hard constraints as before.
The NN has $3$ hidden layers and each layer has $30$ neurons.
Moreover, $100$ points are selected for boundary conditions and $150$ points for initial conditions. 
The training starts with Adam optimizer for $10000$ epochs and 
continues with L-BFGS for $15000$ epochs.

\begin{figure}[htbp]
\centering { \subfigure[Density]{\includegraphics[width=0.48\columnwidth]{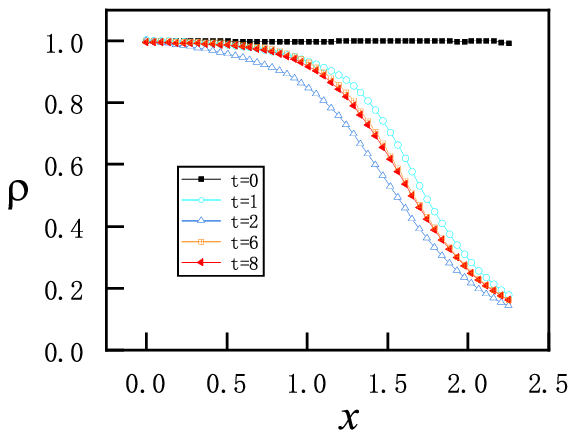}}\quad
\subfigure[Mach number]{\includegraphics[width=0.48\columnwidth]{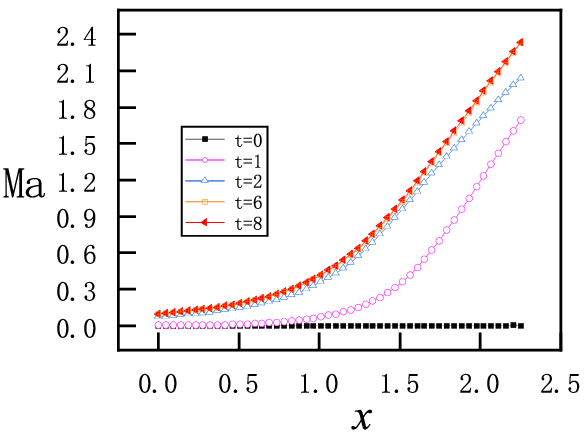}}\quad 
\subfigure[Temperature]{\includegraphics[width=0.48\columnwidth]{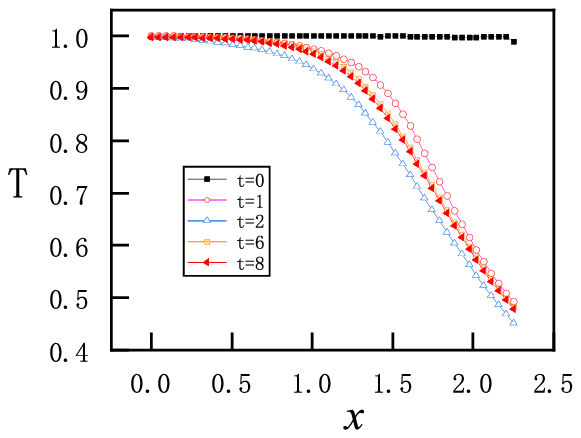}}\subfigure[Pressure]{\includegraphics[width=0.48\columnwidth]{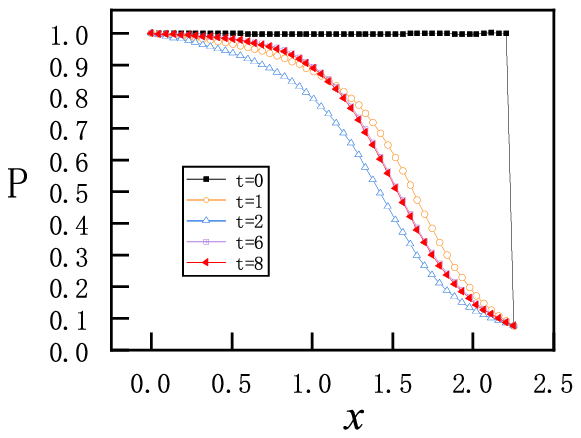}}} \caption{PINNs' solutions for time-dependent subsonic-supersonic flows at $t=0,1,2,3,6,8$. Density, Mach number, temperature and pressure reach steady states for $t\geq 6$.}\label{fig_time012368}
\end{figure}
\begin{figure}[htbp]
\centering \subfigure[Density]{\includegraphics[width=0.48\columnwidth]{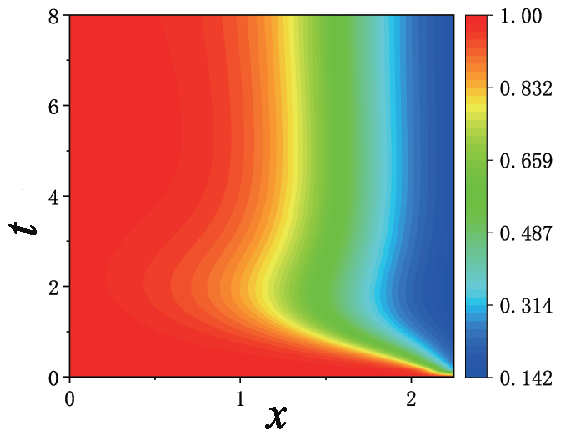}}\quad
{ \subfigure[Mach number]{\includegraphics[width=0.48\columnwidth]{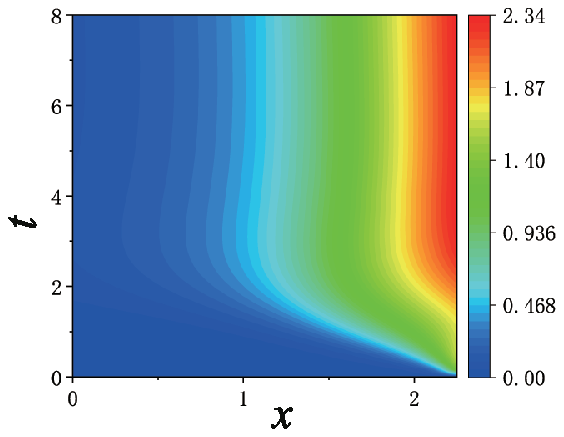}}\quad 
\subfigure[Temperature]{\includegraphics[width=0.48\columnwidth]{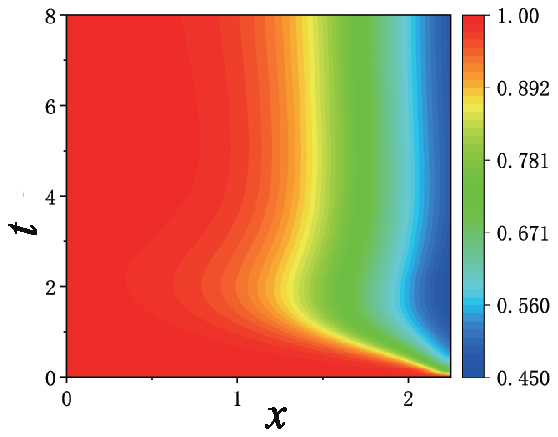}}\quad
\subfigure[Pressure]{\includegraphics[width=0.48\columnwidth]{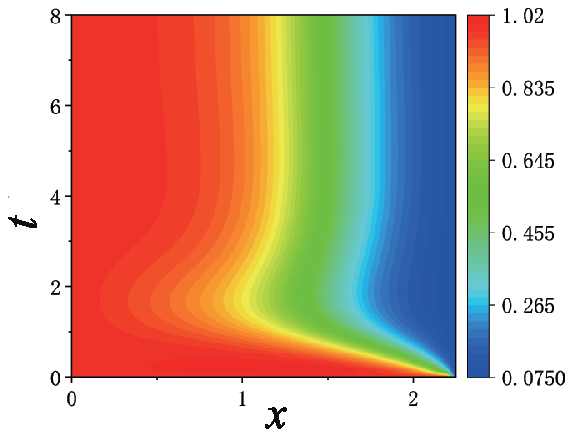}}} \caption{PINNs's solutions for time-dependent subsonic-supersonic flows with continuous evolutions for density, Mach number, temperature and pressure.}\label{fig_fulltime}
\end{figure}

The results of PINNs at $5$ discrete time instants are shown in Fig.~\ref{fig_time012368},
where the flow quickly reaches supersonic state in the diverging part 
and become steady state for $t\geq 6$.
We note that the time evolution for each physical value ($\rho$, $Ma$, $T$, and $P$)
is always monotonic along the $x$ direction.
However, the same evolution of all physical values indicate overshoot in time,
that is, steady profiles of $\rho$, $Ma$, $T$, and $P$ for $t\geq 6$ are in-between profiles at $t=1$ and $t=2$.
Furthermore, we present continuous maps for the physical values in both space and time in Fig.~\ref{fig_fulltime}.
Similarly as in the discrete instants in Fig.~\ref{fig_time012368},
all four physical values in the continuous color maps are monotonic along the $x$ direction,
but overshoot in time before reaching steady states.

\begin{figure}
    \centering
    \includegraphics[width=0.8\columnwidth]{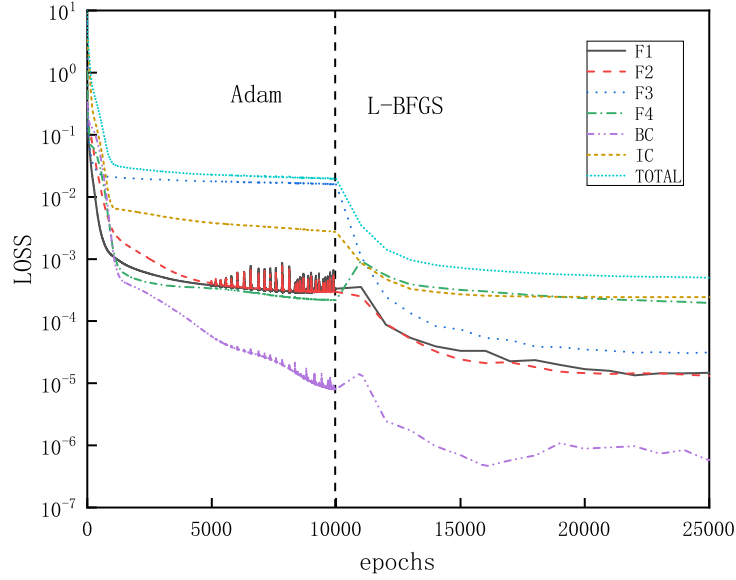}
    \caption{Training loss of PINNs for the continuous time-dependent flows.}
    \label{fig_l3}
\end{figure}
The training loss of PINNs for this problem is shown in Fig.~\ref{fig_l3},
where a catenation of L-BFGS after Adam is effective to reduce the loss
to a much lower level.

\subsection{Exploration of neural networks' parameters for discontinuous flows}\label{sec4.2}
The computational domain is given as $x\in \left [ 0,2.25 \right ]$ and $t\in \left [ 0,25 \right ] $.
The initial and boundary conditions for the time-dependent flow are as follows
\begin{gather*}
    (\rho^{t=0}_{x},v_{x}^{t=0},T_{x}^{t=0},P_{x}^{t=0})=(1.0, 0.0, 1.0, 1.0),\\
(\rho^{t} _{x=0},T_{x=0}^{t},P_{x=0}^{t})=(1.0, 1.0, 1.0), \quad
P_{x=2.25}^{t}=P_{b}=0.81017.
\end{gather*}
The steady state with these boundary conditions was studied in section~\ref{sec3.2},
which corresponds to a flow with a normal shock at $x=1.875$ in the diverging part.
When solving the unsteady process with discontinuity, 
we find that the NNs with the previous setting result in a poor prediction for the solutions at steady state. 
It is understandable, as we have one extra dimension of time for the physical quantities to evolve.
Therefore, we commence to explore the effects of the parameters of NNs and
examine PINNs' solutions at steady states after going through the time-dependent states.

\begin{table}[htbpp]
\centering
\begin{tabular}{|c|c|c|c|}
\hline
\diagbox{index name}{parameters}  & layers $\times$ neurons & regular points  & extra  points \\ \hline
${NN_a}$ & $3 \times 30$   & $100 \times 100$   &  N.A.\\
\hline
${NN_b}$ & $4 \times 50$    & $100 \times 100$   &   N.A. \\  
\hline
${NN_c}$ & $3 \times 30$    &  $100 \times 100$ &  $30 \times 30$   \\ 
\hline
${NN_d}$ & $4 \times 50$    & $100 \times 100$ &  $30 \times 30$  \\ 
\hline 
\end{tabular}
\caption{List of four sets of NNs' parameters.  For initial and boundary conditions $150$ and $100$ points are applied universally. The extra points are within the space-time domain of the diverging part.
Corresponding results are shown in Figs.~\ref{fig_nna} and \ref{fig_nnd} for $NN_a$ and $NN_d$; Figs.~\ref{fig_nnb}, \ref{fig_nnc} in Appendix~\ref{sec_appendix1} for $NN_b$ and $NN_c$.}
\label{table_nn_parameters}
\end{table}

We employ four sets of parameters as listed in Table~\ref{table_nn_parameters},
where we have two architectures of NNs: $3$ hidden layers $\times$ $30$ neurons
and $4$ hidden layers $\times$ $50$ neurons.
For the training points, we have $100 \times 100$ regular points uniformly distributed
in the whole space-time domain $x \times t \in [0,2.25] \times [0, 25] $.
As an attempt to enhance resolution, we add $30 \times 30$ extra points
uniformly distributed in the space-time domain of the diverging region of the nozzle 
$x \times t \in [1.5,2.25] \times [0, 25] $.
For the initial and boundary conditions, 
$150$ and $100$ points are universally applied, respectively.
Each training starts with Adam optimizer for $10000$ epochs and continues with L-BFGS for $15000$ epochs.

\begin{figure}[htbp]
\centering { \subfigure[Density]{\includegraphics[width=0.48\columnwidth]{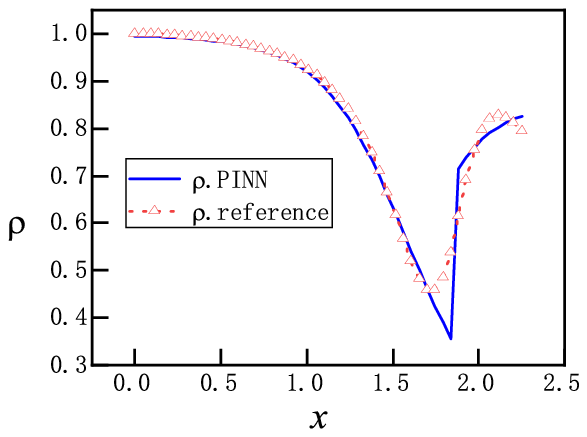}}\quad 
\subfigure[Mach number]{\includegraphics[width=0.48\columnwidth]{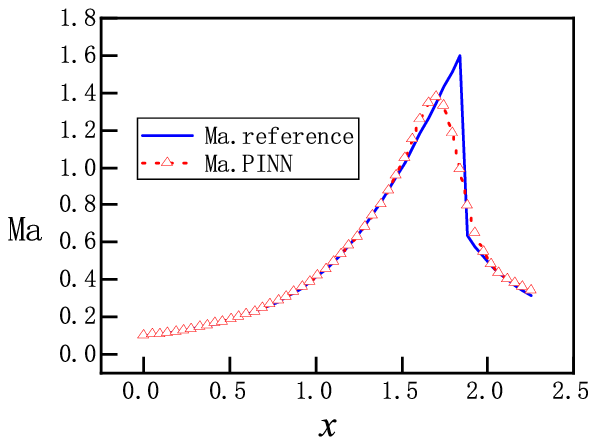}}
\quad
\subfigure[Temperature]{\includegraphics[width=0.48\columnwidth]{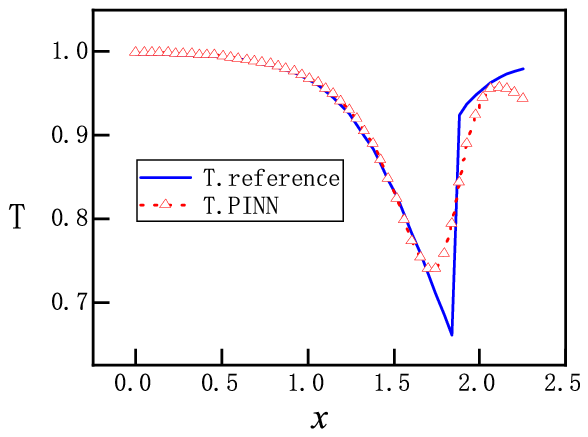}}\subfigure[Pressure]{\includegraphics[width=0.48\columnwidth]{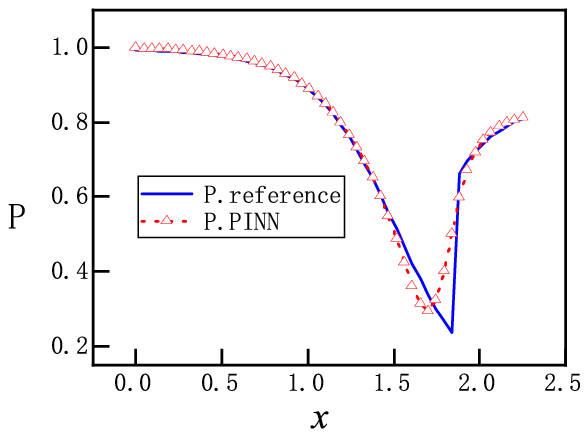}}} \caption{PINNs' results with setup $NN_a$ for steady states from unsteady process: $3$ hidden layers and each layer $30$ neurons; Regular $100 \times 100$ training points for space-time domain $x \times t \in [0, 2.25] \times [0, 25].$}\label{fig_nna}
\end{figure}

\begin{figure}[htbp]
\centering { \subfigure[Density]{\includegraphics[width=0.48\columnwidth]{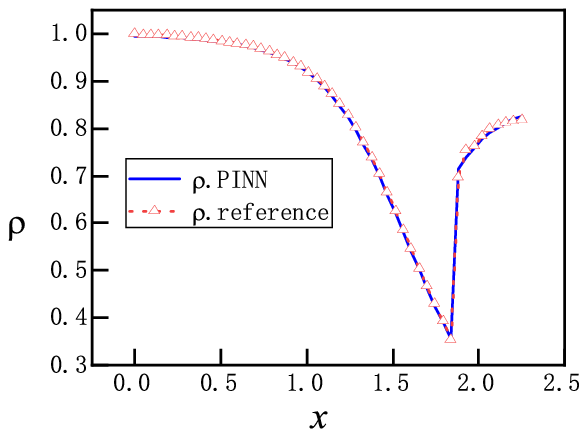}}\quad 
\subfigure[Mach number]{\includegraphics[width=0.48\columnwidth]{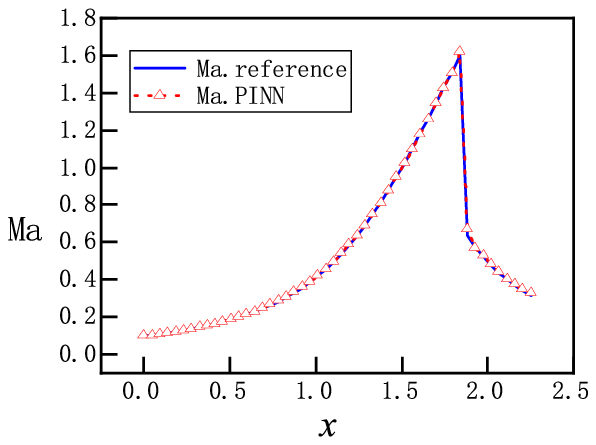}}
\quad
\subfigure[Temperature]{\includegraphics[width=0.48\columnwidth]{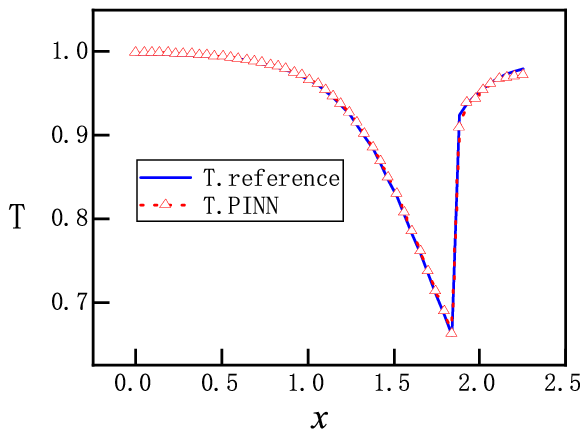}}\subfigure[Pressure]{\includegraphics[width=0.48\columnwidth]{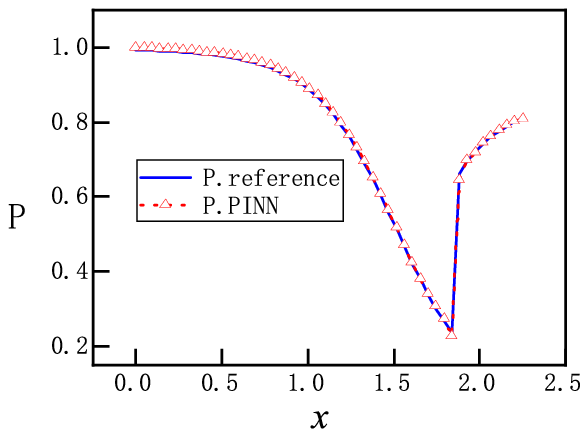}}} \caption{PINNs' results with setup $NN_d$ for steady states from unsteady process: $4$ hidden layers and each layer $50$ neurons; Regular $100 \times 100$ training points for space-time domain $x \times t \in [0, 2.25] \times [0, 25].$
Extra $30 \times 30$ training points for space-time domain $x \times t \in [1.5, 2.25] \times [0, 25]$.
}\label{fig_nnd}
\end{figure}

For the time being, we discard PINNs' results at transit states
and present solutions at steady states.
We observe that the first setup of $NN_a$ reproduces the steady states qualitatively, as shown in Fig.~\ref{fig_nna},
where an norm shock can be identified albeit at a location biased towards upstream.
The overall profiles of $\rho$, $Ma$, $T$ and $P$ are accurate
for the subsonic region before the shock, and deteriorate evidently after the shock.
Next, we consider two individual improvements over the NNs:
one is with enhanced number of layers and neurons corresponding to setup $NN_b$;
another is with enlarged number of sampling points in the diverging region of the nozzle
corresponding to setup $NN_c$.
Both setups improve the results substantially, but the solutions are still not sufficiently accurate, as shown in Figs.~\ref{fig_nnb} and \ref{fig_nnc} in Appendix~\ref{sec_appendix1}.
Furthermore, we look at PINNs' results with setup $NN_d$ in Fig.~\ref{fig_nnd},
which represents the combined efforts of improvement on the NN's achitecture
and increased number of training points.
We observe that solutions from PINNs solving an unsteady process
predict the steady states correctly,
with the same accuracy as the solutions of a steady flow solved by PINNs.

\begin{figure}[htbp]
\centering {\subfigure[Density]{\includegraphics[width=0.48\columnwidth]{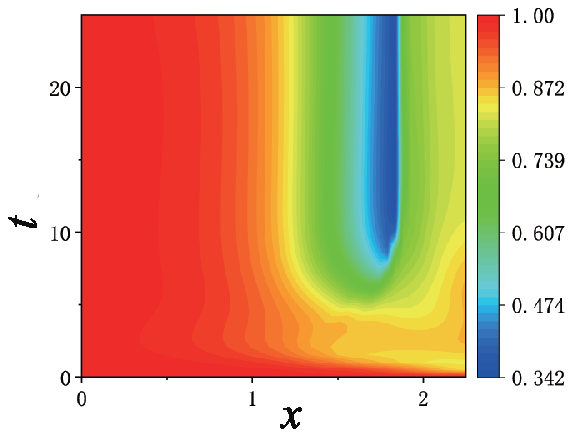}} \quad 
\subfigure[Mach number]{\includegraphics[width=0.48\columnwidth]{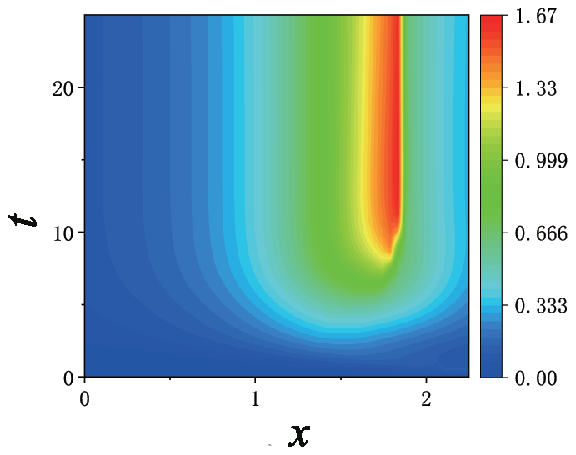}}
\quad
\subfigure[Temperature]{\includegraphics[width=0.48\columnwidth]{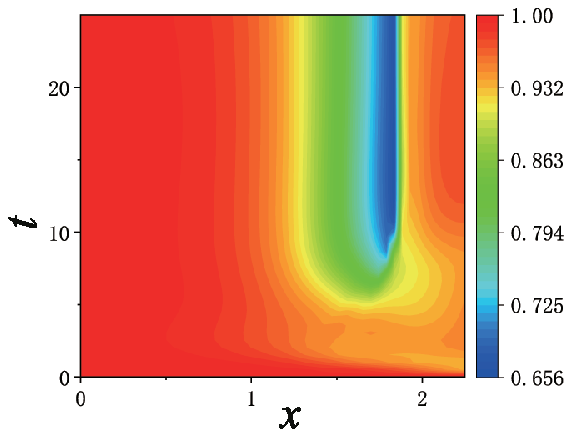}}\subfigure[Pressure]{\includegraphics[width=0.48\columnwidth]{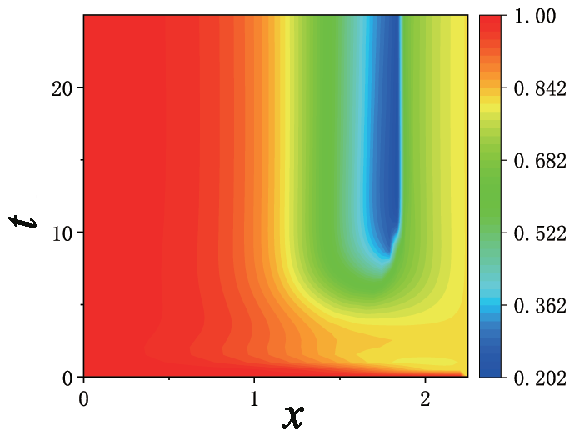}}} \caption{PINNs's results for time-dependent subsonic-supersonic flows with discontinuous evolutions for density, Mach number, temperature and pressure. The setup of PINNs is with $NN_{d}$ as listed in Table~\ref{table_nn_parameters}.}
\label{fig_nnd_fulltime}
\end{figure}
Lastly, we present the complete evolution of $\rho$, $Ma$, $T$ and $P$ in Fig.~\ref{fig_nnd_fulltime}.
From the color maps of physical values in the whole space-time domain,
we may divide approximately the evolution into three stages in time.
At the first stage for $t \lesssim 5$, the flow has clearly subsonic characteristics:
all four physics values develop rapidly, but always have continuous profiles.
At the second stage between $5 \lesssim t \lesssim 8.5$,
supersonic characteristics of the flow arise at downstream of the throat at $x=1.5$,
but the physical values are still continuous.
At the third stage for $ t \gtrsim 8.5$,
discontinuous phenomena emerge.
Later on, physical values settle for $t \gtrsim 10$ ,
where a sharp interface for physics values is evident at $x=1.875$.

\subsection{Two initial conditions for discontinuous flows}\label{sec4.3}
The transient flows may be very different under distinct initial conditions.
Therefore, we consider the effects due to two initial conditions.  
The first initial condition as before is repeated as follows
\begin{gather*}
    (\rho^{t=0}_{x},v_{x}^{t=0},T_{x}^{t=0},P_{x}^{t=0})=(1.0, 0.0, 1.0, 1.0).
\end{gather*}
This corresponds to the scenario when the inlet is already open 
and the outlet is closed before the flow.
Therefore, the density and pressure inside the nozzle are identical as the stagnation values of the inlet.
As second initial condition, we consider that the outlet is open and the inlet is closed before the flow.
Therefore, the density and pressure inside the nozzle are identical as the values of the outlet.
The values for the second initial condition are as follows
\begin{gather*}
    (\rho^{t=0}_{x},v_{x}^{t=0},T_{x}^{t=0},P_{x}^{t=0})=({\rho}_{b}=0.81017, 0.0, 1.0, P_{b}=0.81017).
\end{gather*}
For both cases, the temperature initially has a stagnation value of $T=1.0$ in the entire nozzle and the boundary conditions are identical as follows
\begin{gather*}
(\rho^{t} _{x=0},T_{x=0}^{t} ,P_{x=0}^{t})=(1.0, 1.0, 1.0), \quad
P_{x=2.25}^{t}=P_{b}=0.81017.
\end{gather*}

\begin{figure}[htbp]
\centering {\subfigure[Density]{\includegraphics[width=0.48\columnwidth]{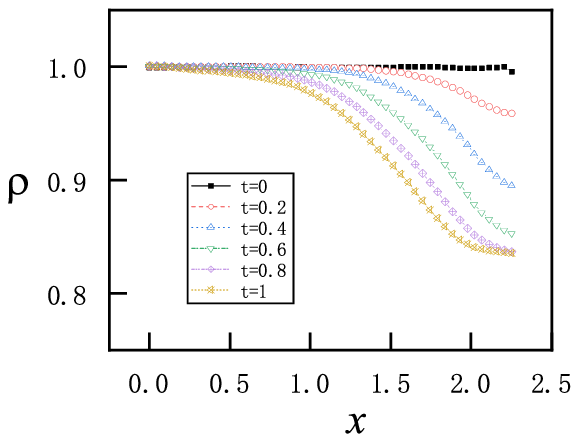}} \quad 
\subfigure[Mach number]{\includegraphics[width=0.48\columnwidth]{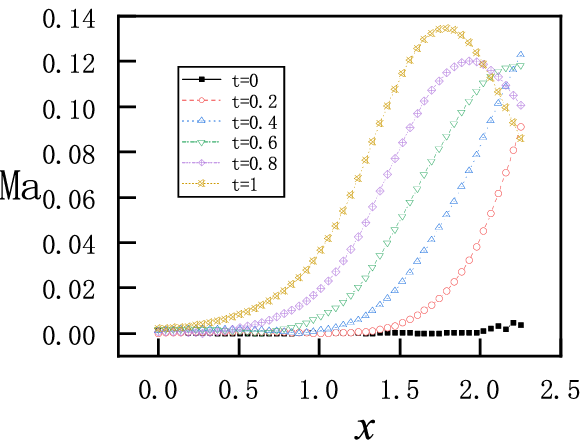}}
\quad
\subfigure[Temperature]{\includegraphics[width=0.48\columnwidth]{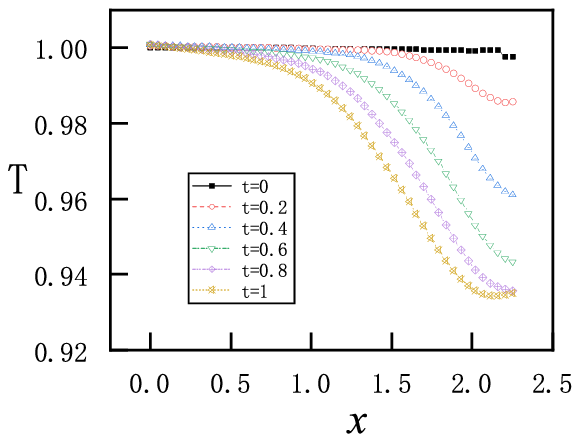}}\subfigure[Pressure]{\includegraphics[width=0.48\columnwidth]{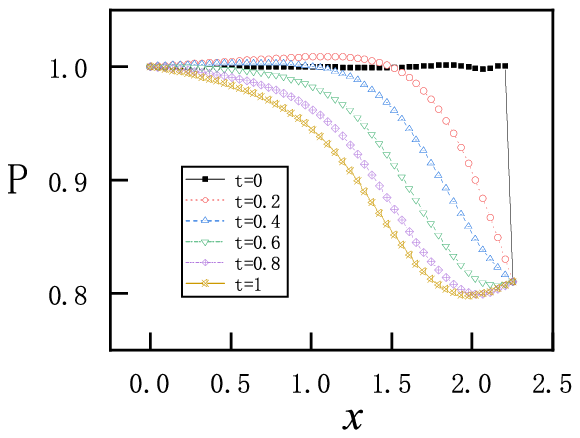}}} \caption{PINNs' results at $6$ instants of short time for time-dependent discontinuous flows with the first initial conditions: before the flow the inlet is open while the outlet is closed.}\label{fig_under_stagnation_pressure0-1}
\end{figure}

\begin{figure}[htbp]
\centering { \subfigure[Density]{\includegraphics[width=0.48\columnwidth]{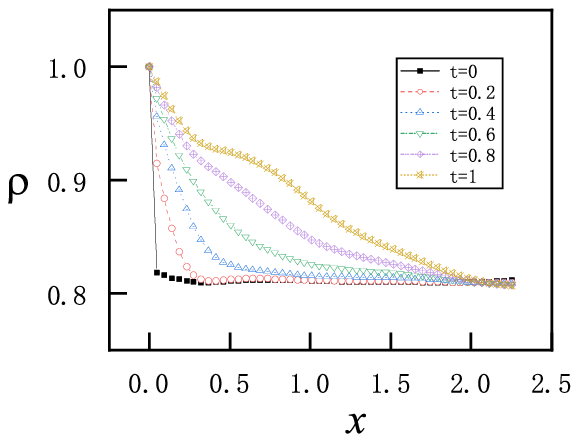}}\quad 
\subfigure[Mach number]{\includegraphics[width=0.48\columnwidth]{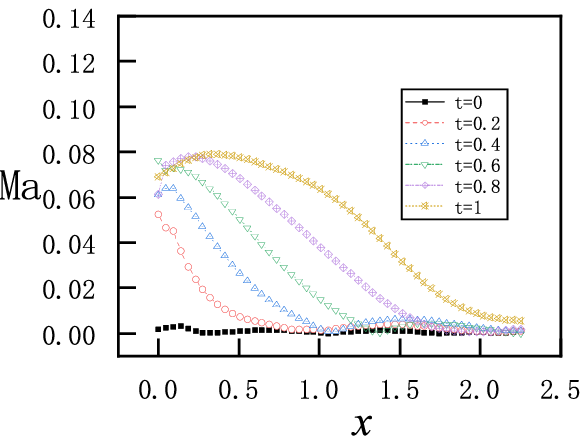}}
\quad
\subfigure[Temperature]{\includegraphics[width=0.48\columnwidth]{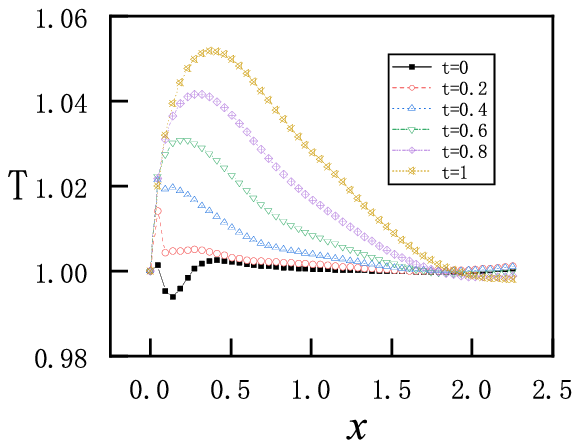}}\subfigure[Pressure]{\includegraphics[width=0.48\columnwidth]{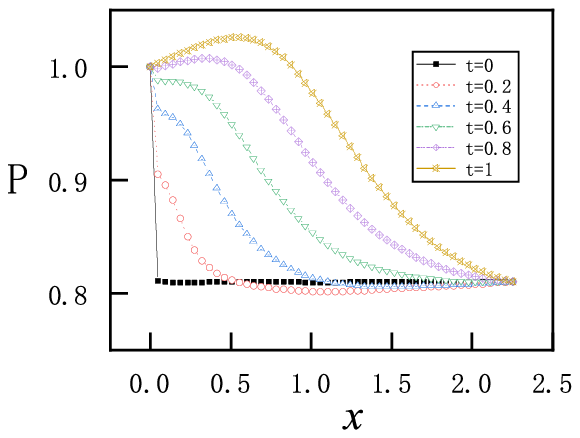}}} \caption{PINNs' results at $6$ instants of short time for time-dependent discontinuous flows with the second initial conditions: before the flow the inlet is closed while the outlet is open.}\label{fig_under_atmospheric_pressure0-1}
\end{figure}

We observe that both flows develop rapidly from the two different initial conditions.
After $t=1$, the two paths of transition states of all physical values are already very similar, 
as shown in Figs.~\ref{fig_under_stagnation_pressure0-25} and \ref{fig_under_atmospheric_pressure0-25} in Appendix~\ref{sec_appendix2}.
Therefore, here we only present the two sets of results for $t \in [0, 1]$ in Figs.~\ref{fig_under_stagnation_pressure0-1} and \ref{fig_under_atmospheric_pressure0-1}. 
With the first initial conditions, $\rho$, $v$ and $T$ evolve quickly but smoothly 
except for $P$, as there is a discontinuous jump between the boundary value and inner initial values at the outlet.
Nevertheless, the pressure becomes continuous after $t=0.2$, as shown in Fig.~\ref{fig_under_stagnation_pressure0-1}(d).
With the second initial conditions, $v$ and $T$ values evolve quickly but smoothly
except for $\rho$ and $P$, as there are discontinuous jumps between the boundary value and inner initial values at the inlet.
Nevertheless, both the density and pressure become continuous after $t=0.2$, as shown in Figs.~\ref{fig_under_atmospheric_pressure0-1}(a) and (d).

\section{Solutions in conservative form}\label{sec5}
In classical numerical methods a conservative form of the PDEs is in favor and
numerical solutions of the non-conservative form tend to be unstable. 
Therefore, we consider the solution procedure of PINNs in the context of conservative form,
which is given as follows
\begin{align}
\partial _t U + \nabla  \cdot K = J.
\end{align}
$U$, $K$ and $J$ are vectors:
$U=\begin{pmatrix}
U_1 \\
U_2\\
U_3
\end{pmatrix},$
$K=\begin{pmatrix}
K_1 \\
K_2\\
K_3
\end{pmatrix}$
$J=\begin{pmatrix}
J_1 \\
J_2\\
J_3
\end{pmatrix}$,
and they are defined as
\begin{align}
 \label{U1}
 \left \{\aligned
\rho A = U_1,\\
\rho Av = U_2,\\
\rho A(\frac{T}{\gamma -1}  +\frac{\gamma }{2} v^{2} ) = U_3.
 \endaligned
 \right.
 \end{align} 
 
\begin{align}
 \label{K}
 \left \{\aligned
\rho Av = K_1,\\
\rho Av^2 + \frac{1}{\gamma }  PA= K_2,\\
\rho A(\frac{T}{\gamma -1}  +\frac{\gamma }{2} v^{2} ) +PAv  = K_3.
 \endaligned
 \right.
 \end{align} 

 \begin{align}
 \label{J}
 \left \{\aligned
0 = J_1,\\
\frac{1}{\gamma }  P\frac{\partial A}{\partial x}  = J_2,\\
0  = J_3.
 \endaligned
 \right.
 \end{align} 

\begin{figure}
\centering
\includegraphics[width=100mm]{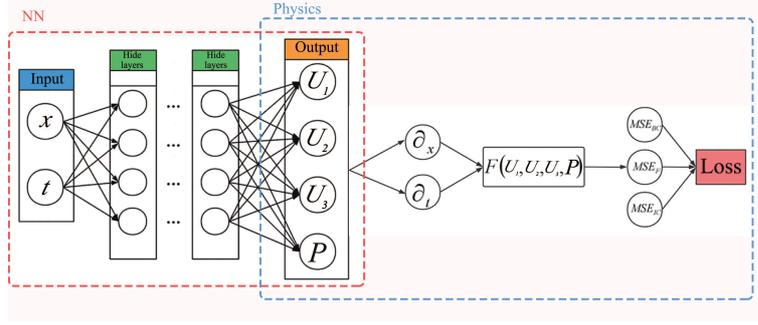}
\caption{The structure of PINNs for Euler equations in conservative form.}
 \label{nn_for_conservation}
\end{figure}
Accordingly, we have to adjust the outputs of the NNs to be $U_1,U_2,U_3 $ and $P$, 
as shown in Fig.~\ref{nn_for_conservation}.
To facilitate the construction of loss function, 
we have to rewrite $K_i$ as a function of the NNs' outputs, namely, $U_i$, as follows
\begin{align}
 \left \{\aligned
U_2 = K_1,\\
\frac{U_{2}^{2} }{U_1} +\frac{\gamma -1}{\gamma } ( U_3-\frac{\gamma }{2}\frac{U_{2}^{2} }{U_1} ) = K_2,\\
\gamma \frac{U_{2}U_{3}}{U_1} -\frac{\gamma (\gamma -1)}{2} \frac{U_{2}^{3} }{U_{1}^{2} }  = K_3,
 \endaligned
 \right.
 \label{eq_ks}
 \end{align}
where we have used the fact $P=\rho T$ from the equation of state.
Therefore, each component of the loss function is as follows
\begin{align}
 \left \{\aligned
\frac{\partial U_{1}  }{\partial t} +\frac{\partial K_{1} }{\partial x} =F_1(x,t),\\
\frac{\partial U_{2}  }{\partial t} +\frac{\partial K_{2} }{\partial x}-\frac{1}{\gamma}P\frac{\partial A}{\partial x}  =F_2(x,t),\\
\frac{\partial U_{3}  }{\partial t} +\frac{\partial K_{3} }{\partial x} =F_3(x,t),\\
P-\frac{U_{1} }{A} \left ( \gamma -1 \right ) \left ( \frac{U_3}{U_1}-\frac{\gamma }{2} \frac{U_{2}^{2} }{U_{1}^{2} }   \right )=F_4(x,t).
 \endaligned
 \right. 
 \label{eq_loss_new}
 \end{align} 
Here $F_1(x,t)$, $F_2(x,t)$, $F_3(x,t)$, and  $F_4(x,t)$ represent residuals of the mass, momentum, energy, and state equations, respectively.
For automatic differentiation, we have to expand all terms $K_i$ in Eqs.~(\ref{eq_loss_new}) with $U_i$ from Eqs.~(\ref{eq_ks}).
These expansions quickly become unpleasant with many dividing operations,
which poses challenges for gradient calculations and optimizations of the NNs' parameters.
Despite our best efforts, the loss function does not descend easily
and no meaningful predictions are made by PINNs with the conservative form.

\section{Conclusions}
\label{sec6}
We applied physics-informed neural networks or PINNs to directly solve steady and time-dependent compressible flows within a converging-diverging channel, corresponding to an unsupervised learning.
With different boundary conditions, the flow may be completely subsonic, 
subsonic via a smooth transition to supersonic,
or further from supersonic via a discontinuous transition back to subsonic.

Firstly, for a simple diverging channel,
when sonic boundary conditions are imposed at the inlet and the outlet are left intentionally free,
PINNs provide a subsonic or a supersonic continuous solution randomly and deliberately avoid solutions with discontinuity. 
The stochastic outcomes result from the randomness during initialization of the neural networks (NNs);
Secondly, for a converging-diverging nozzle,
PINNs with a default setting cannot capture transition from supersonic to subsonic flows
with a discontinuity for the normal shock,
although given proper boundary conditions at both the inlet and outlet.
Instead, the loss function is unwilling to descend during training and 
consequently, a trivial solution with zero velocity and incorrect continuous profiles for other physical values are obtained.
The two examples above indicate that during training the optimizer somehow
``minimizes its efforts" before minimizing the loss function:
it is inclined to offer smooth solutions (right or wrong) and 
reluctant to find discontinuous solutions.
Hence, a small remedy of PINNs is pertinent.

After a close inspection on the descent of each component of the loss function and predictions of the physical values,
we promote to put $20$ times more weight on the minimization of the momentum equation
and meanwhile enforce hard constraints on the boundary conditions of pressure.
This exertion is coincident with a recent effort put forward by Perdikaris' group~\supercite{Wang2021},
where a dynamic weight is proposed to balance the gradients among different 
components of the loss function, to mitigate gradient vanishing.
For the problems considered in this work, 
we acknowledge a constant heavy weight for the momentum loss being adequate for accurate solutions,
after a couple of numerical experiments with trial and error.

With $90$ neurons and $100$ training points,
the so-modified version of PINNs is able to deliver accurate solutions at steady states for subsonic flows, supersonic flows and mixture of both with a norm shock as sharp transition from supersonic to subsonic flows.
For unsteady processes, with $200$ neurons and $16900$ training points,
PINNs are able to predict accurately the time-dependent flows until steady states.
Whether for steady or unsteady flows, PINNs are able to 
identify the locations of shocks accurately and offer very sharp profiles for the transitions without Gibbs phenomenon.
These results are promising, as they encourage us to apply
PINNs to solve more discontinuous physics phenomena
and to replace/supplement traditional numerical schemes to a certain extent.
Finally, when the PDEs are expressed in the conservative form, 
which is in favor by traditional numerical schemes,
the output terms of the NNs and their corresponding loss function
are entangled and not viable for an effective optimization.
Consequently, the predictions offered by PINNs are incorrect.
This indicates that PINNs prefer the simple differential form of the PDEs
over the conservative form, as the former is more appropriate 
for straightforward automatic differentiation during optimization.

We envisage two research lines beyond this work. One is to explore PINNs as a direct numerical solver for more general compressible flows in two and three dimensions, especially with shock phenomena.
From the performance on one-dimensional flows, 
it seems promising for PINNs to solve more compressible/discontinuous flows,
where no exquisite shock-capturing schemes are essential.
Another one is to supply PINNs with partially available experimental data,
such as a few pressure values via sensors within the flow,
to recover other physics values
and/or to estimate unknown coefficients in the PDEs,
such as wall frictions.
\\
\\
\section*{Acknowledgments}
X. Bian received the starting grant from 100 talents program of Zhejiang University.
This work is partially supported by Hangzhou Shiguangji Intelligient Electronics Technology Co., Ltd, Hangzhou, China.
The authors appreciate discussions with Dr. Bonan Xu and Mr. Yongzheng Zhu.

\bibliographystyle{unsrt}
\bibliography{ref}

\renewcommand{\theequation}{A\arabic{equation}}
\setcounter{equation}{0}

\footnotesize \noindent\textbf{Appendix}
\\
\appendix
\section{Different neural networks' parameters for discontinuous flows}\label{sec_appendix1}
\begin{figure}[htbp]
\centering { \subfigure[Density]{\includegraphics[width=0.48\columnwidth]{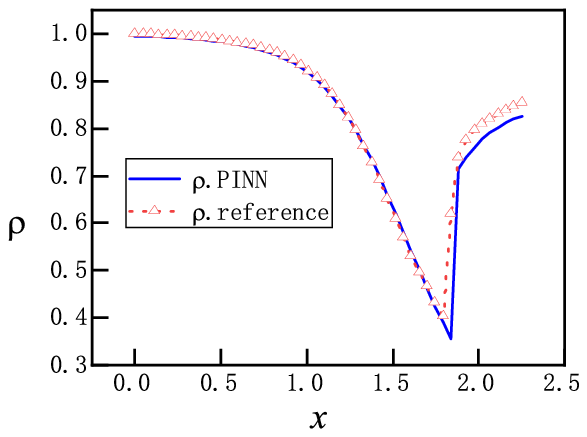}}\quad 
\subfigure[Mach number]{\includegraphics[width=0.48\columnwidth]{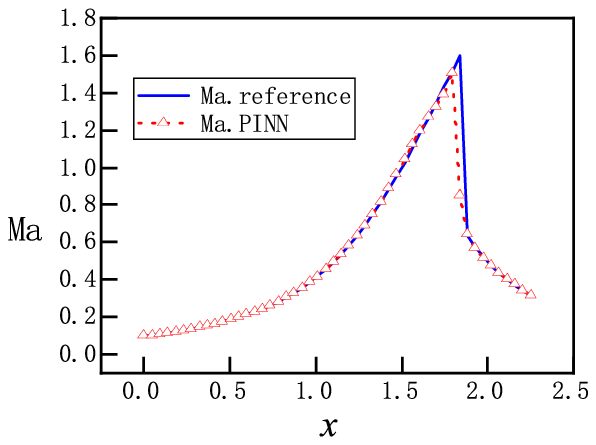}}
\quad
\subfigure[Temperature]{\includegraphics[width=0.48\columnwidth]{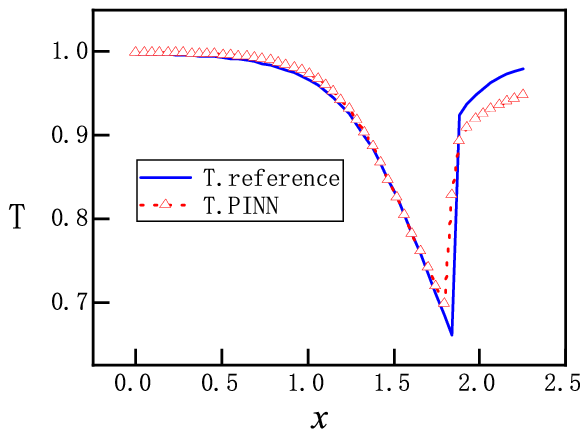}}\subfigure[Pressure]{\includegraphics[width=0.48\columnwidth]{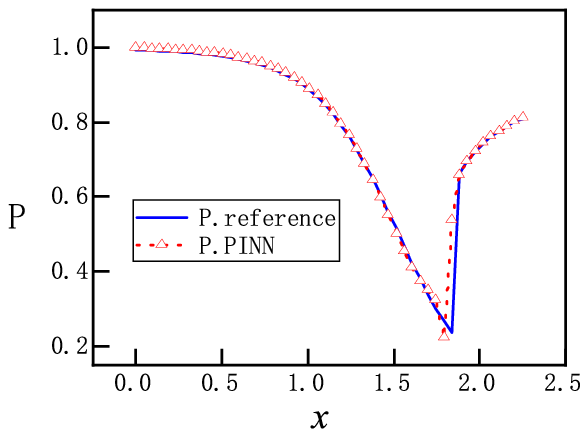}}} \caption{PINNs' results with setup $NN_b$ for steady states from unsteady process: $4$ hidden layers and each layer $50$ neurons; Regular $100 \times 100$ training points for space-time domain $x \times t \in [0, 2.25] \times [0, 25].$}\label{fig_nnb}
\end{figure}

\begin{figure}[htbp]
\centering { \subfigure[Density]{\includegraphics[width=0.48\columnwidth]{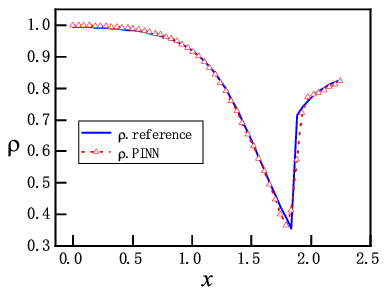}}\quad 
\subfigure[Mach number]{\includegraphics[width=0.48\columnwidth]{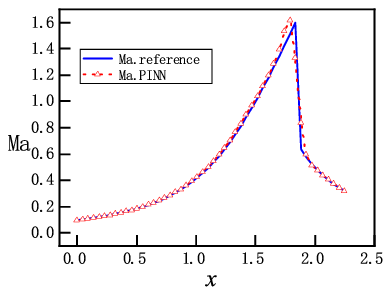}}
\quad
\subfigure[Temperature]{\includegraphics[width=0.48\columnwidth]{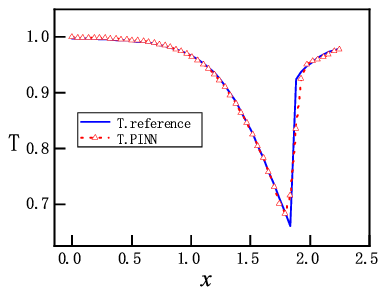}}\subfigure[Pressure]{\includegraphics[width=0.48\columnwidth]{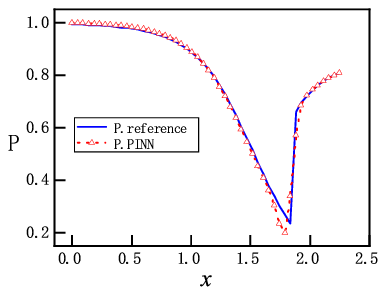}}} \caption{PINNs' results with setup $NN_c$ for steady states from unsteady process: $3$ hidden layers and each layer $30$ neurons; Regular $100 \times 100$ training points for space-time domain $x \times t \in [0, 2.25] \times [0, 25].$
Extra $30 \times 30$ training points for space-time domain $x \times t \in [1.5, 2.25] \times [0, 25].$
}\label{fig_nnc}
\end{figure}

Figs.~\ref{fig_nnb} and \ref{fig_nnc} present the results of the NNs under the two settings of $NN_b$ and $NN_c$ in Table~\ref{table_nn_parameters} of Sec.~\ref{sec4.2}.
Either increasing the number of freedoms of NNs or adding more training points improve predictions to a certain extent, but the results are not sufficiently accurate.
More specifically, there are still mismatches for physics values near the shock location
from both $NN_b$ and $NN_c$ and profiles of $\rho$ and $T$ from $NN_b$
clearly misalign with analytical references after the shock.

\section{Two initial conditions for time-dependent discontinuous flows}\label{sec_appendix2}
The following two diagrams present flows from time $0$ to $25$ with two different initial conditions in Sec.~\ref{sec4.2}. As observed from the two sets of plots, flows do not differentiate from each other after $t=1$ between the two initial conditions. 
These results suggest that one should examine the flows for $t \in [0, 1]$ to compare the differences.
\begin{figure}[htbp]
\centering { \subfigure[Density]{\includegraphics[width=0.48\columnwidth]{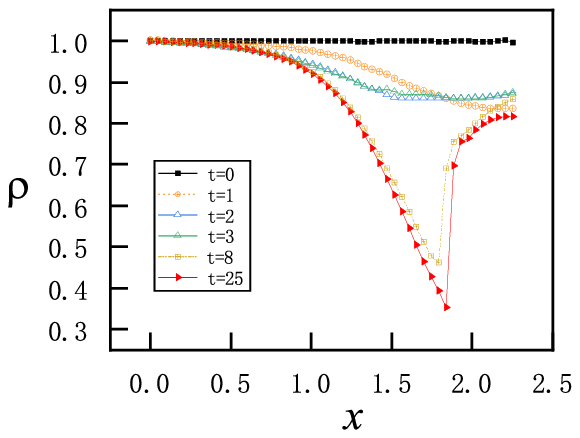}}\quad 
\subfigure[Mach number]{\includegraphics[width=0.48\columnwidth]{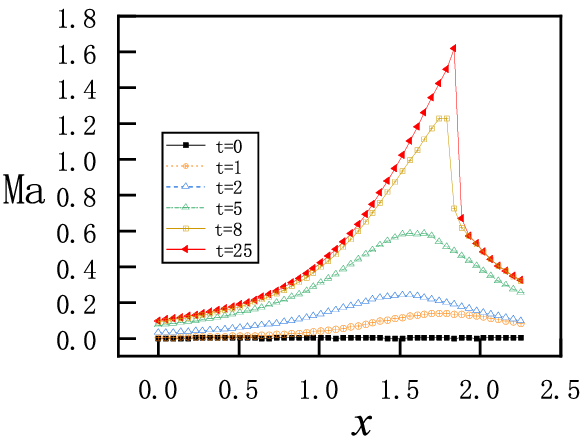}}
\quad
\subfigure[Temperature]{\includegraphics[width=0.48\columnwidth]{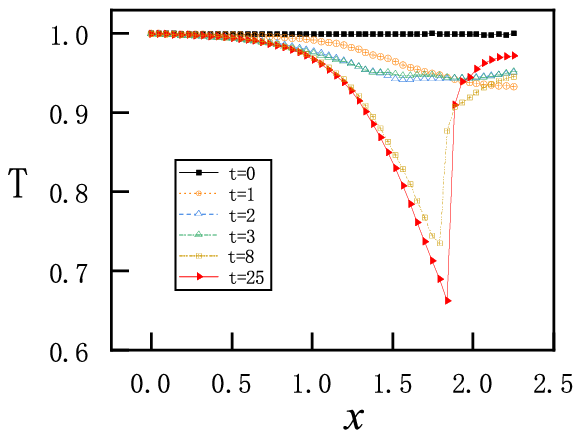}}\subfigure[Pressure]{\includegraphics[width=0.48\columnwidth]{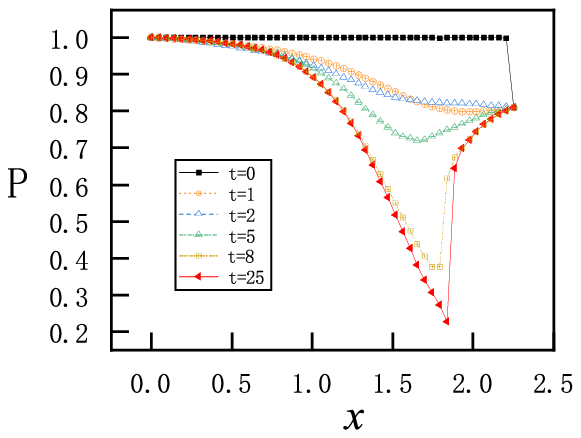}}} \caption{PINNs' results at $6$ instants of long time for time-dependent discontinuous flows with the first initial conditions: before the flow the inlet is open while the outlet is closed.}\label{fig_under_stagnation_pressure0-25}
\end{figure}

\begin{figure}[htbp]
\centering { \subfigure[Density]{\includegraphics[width=0.48\columnwidth]{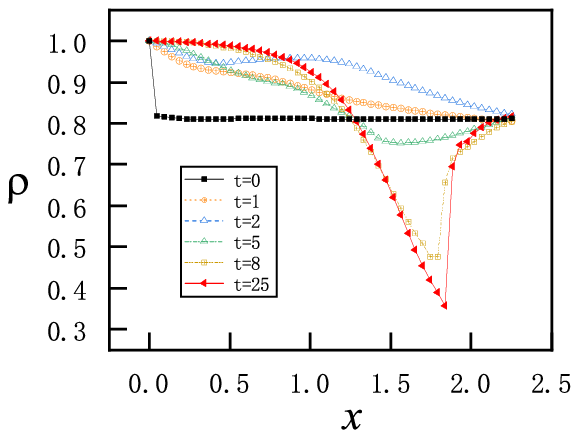}}\quad
\subfigure[Mach number]{\includegraphics[width=0.48\columnwidth]{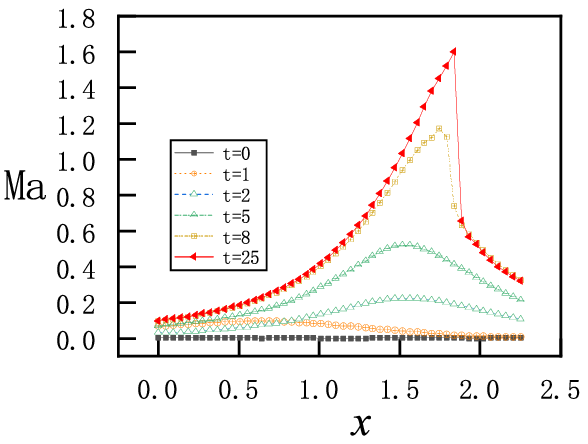}}
\quad
\subfigure[Temperature]{\includegraphics[width=0.48\columnwidth]{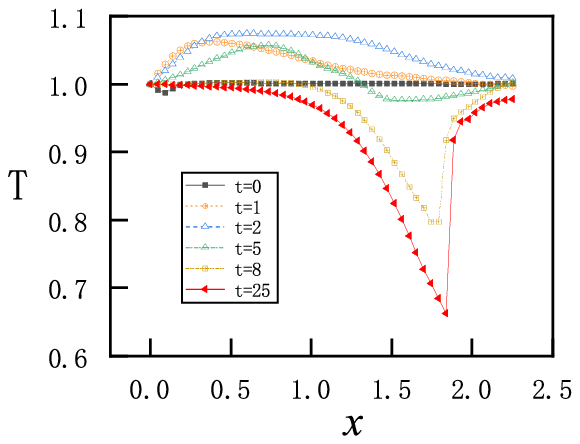}}\subfigure[Pressure]{\includegraphics[width=0.48\columnwidth]{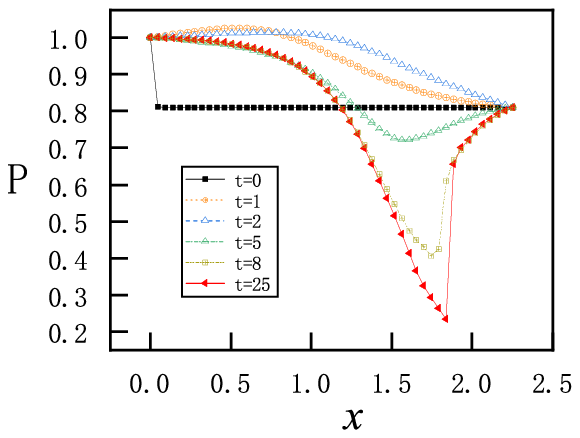}}} \caption{PINNs' results at $6$ instants of long time for time-dependent discontinuous flows with the second initial conditions: before the flow the inlet is closed while the outlet is open.}\label{fig_under_atmospheric_pressure0-25}
\end{figure}

\end{CJK*}
\end{document}